\documentclass[aps,prb,twocolumn, groupaddress,floatfix]{revtex4}
\usepackage{graphicx}
\usepackage{epstopdf}
\usepackage{amssymb}
\usepackage{epsf}
\newcommand{\bra}[1]{\langle #1|}
\newcommand{\ket}[1]{|#1\rangle}
\newcommand{\braket}[1]{\langle#1\rangle}
\newcommand{\doublebra}[1]{\langle \langle #1|}
\newcommand{\doubleket}[1]{|#1\rangle \rangle}
\newcommand{\doublebraket}[1]{\langle \langle#1\rangle \rangle}

\renewcommand{\phi}{\varphi}
\renewcommand{\epsilon}{\varepsilon}

\renewcommand{\vec}[1]{{\bf #1}}

\newcommand{\eqnref}[1]{Eq.~(\ref{#1})}	
\newcommand{\fref}[1]{Fig.~\ref{#1}}	
\newcommand{\secref}[1]{Section~\ref{#1}}

\newcommand{\beq}{\begin{equation}}
\newcommand{\eeq}{\end{equation}}
\newcommand{\ba}{\begin{array}{ccc}}
\newcommand{\ea}{\end{array}}

\usepackage{amssymb}
\usepackage{amsmath}

\begin{document}
\title{Transport properties of non-equilibrium systems under the application of light: 
Photo-induced quantum Hall insulators without Landau levels}
\author{Takuya Kitagawa$^{1}$ }
\author{Takashi Oka$^{1,2}$ }
\author{Arne Brataas$^{1,3}$}
\author{Liang Fu$^{1}$}
\author{Eugene Demler$^{1}$}

\affiliation{$^{1}$Physics Department, Harvard University, Cambridge,
Massachusetts 02138, USA, $^{2}$
Department of Physics, Faculty of Science, University of Tokyo, Tokyo 113-0033, Japan
,  $^{3}$ Department of Physics, Norweigian University of Science and Technology, NO-7491 Trondheim, Norway}

\date{\today}
\begin{abstract}

In this paper, we study transport properties of non-equilibrium systems 
under the application of light in many-terminal measurements, using the Floquet picture. 
We propose and demonstrate that the quantum transport properties can be controlled 
in materials such as graphene and topological insulators, via the application of light. 
Remarkably, under the application of off-resonant light, topological transport properties can be induced; 
these systems exhibits quantum Hall effects in the absence of a magnetic field with a near quantization of the Hall conductance, realizing so-called quantum Hall systems without Landau levels first proposed by Haldane. 

\end{abstract}

\maketitle
%
%


\section{Introduction}
Application of light is a powerful method to change material properties. For example, light can induce currents through mechanisms such as photovoltaic effect\cite{Glass1974}, photo-thermoelectric effect\cite{Xu2010}, and photo-drag effects\cite{Karch2010}. Moreover, light can change the response of materials and induce insulator to metal transitions\cite{Miyano1997} or change the characteristics of $p-n$ junctions\cite{Syzranov2008}.

In recent years, there has been tremendous developments and interests in the induction of quantum phases 
through light applications. For example,
 experiments have demonstrated that superconductivity can be induced through infrared pulses
in high-temperature cuprate superconductors\cite{fausti2011}. 
Inductions of quantum phases are inherently non-equilibrium phenomena, and thus their understanding 
is quite challenging. Even some basic questions such as the physical signatures of the induced phases 
and how such phases can be stabilized in a steady state do not have answers yet. 
Many of quantum phases manifest themselves through transport, and therefore, understanding of transport properties
in non-equilibrium, open systems is crucial for experimental verifications of such induction of quantum phases 
under the application of light. 
In this paper, we develop a general formalism for studying non-equilibrium transport under the application of light, and, using the formalism, address the possibilities of the induction of topological properties through light.

Motivated by recent rapid development of the understanding in topological phases, the possibility of 
inducing topological phases such as integer quantum Hall phase and topological insulators through light
has been theoretically explored by many different groups\cite{Oka,Oka2009,Kitagawa2010a}. 
Generally speaking, the application of light on electron systems has two important physical effects;
1. photon-dressing of band structures through the mixing of different bands 
2. redistribution of electron occupation numbers through the absorptions/emissions of photons 
leading to non-equilibrium distributions. 
Previous works proposed optical induction of band structures
with topological properties \cite{Oka, Oka2009,Kitagawa2010a}, and thus have mostly focused on the analysis of 
the first effect. On the other hand, most of these works do not address the question 
of the second effect, the redistribution of electrons in the band structure, and thus its physics is yet poorly understood. 
Topological properties only appear when certain bands are fully filled, and it is
not clear how this band occupation can be achieved and topological properties survive
when the system is strongly driven out of equilibrium by the application of light. 


\begin{figure}[t!]
\begin{center}
\includegraphics[width = 8.5cm]{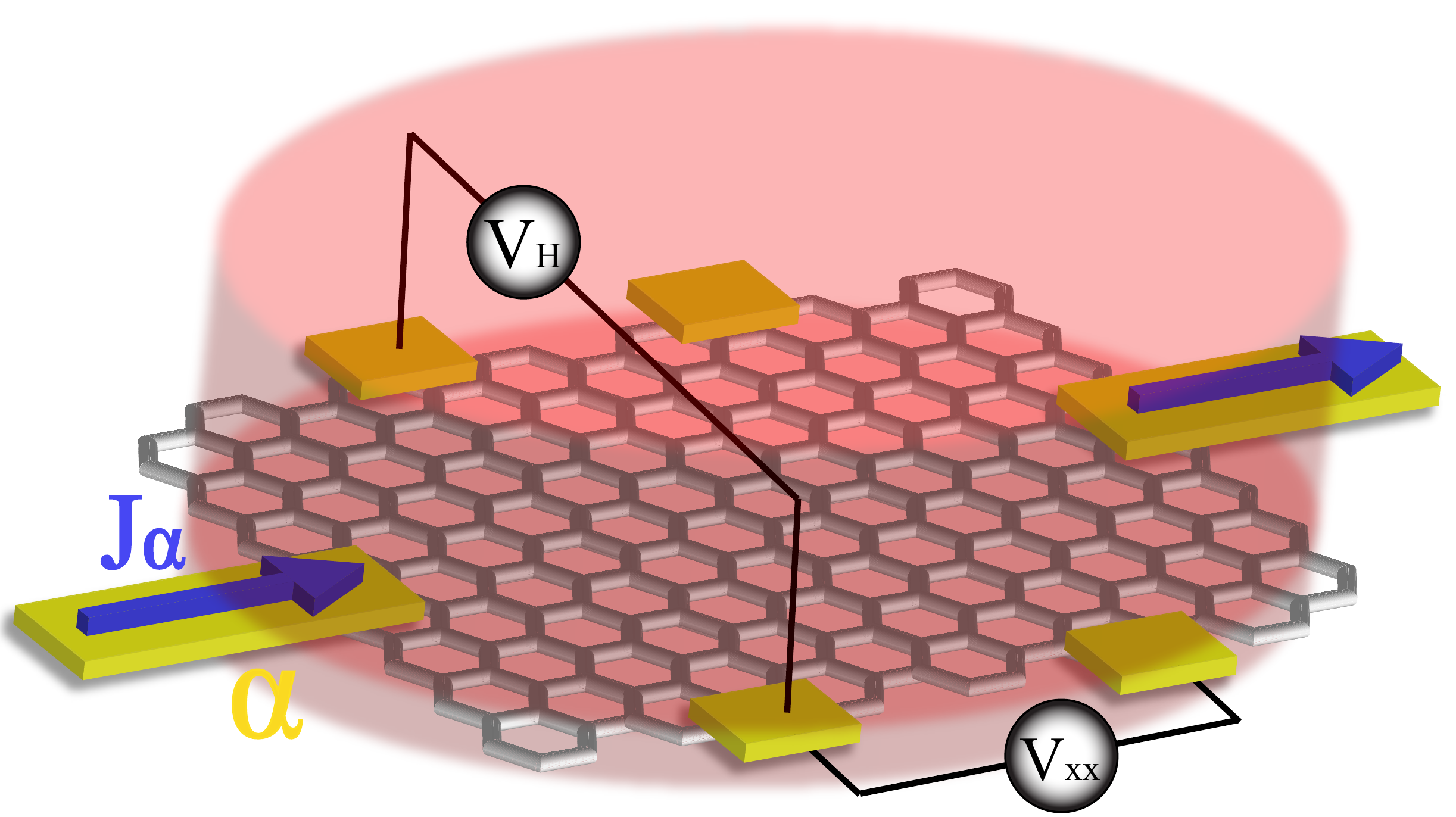}
\caption{The many-terminal measurements of DC current for graphene 
under the application of light. 
Graphene is attached to multiple leads labeled by $\{\alpha\}$, 
and the leads are connected to reservoirs at chemical potentials, $\{ \mu_{\alpha} \}$.
Off-resonant, circularly polarized light is applied to the graphene. 
In the absence of impurities and interactions, electrons coming from leads coherently propagate 
under the application of light, asborb or emit photons, and leak out into leads. 
 Current measurements between each leads determine longitudinal and Hall conductances. 
}
\label{figure1}
\end{center}
\end{figure}

In order to answer these questions, we study the physical consequence of the application of light through DC many-terminal transport measurements as in \fref{figure1}.  The coupling of the 
driven systems with leads, that are in return coupled with equilibrium reservoirs, 
plays the crucial role to determine the occupations of electrons. 
Using the formalism for transport properties in periodically driven systems developed by various groups\cite{Jauho1994},
here we study topological transport phenomena in materials such as graphene and three dimensional topological
insulators. 
First of all, we show that non-equilibrium transport properties cannot generally be captured by the photon-dressed,
effective band structures.
In particular, in addition to the usual transport through such static effective band structures, there are 
contributions from photon assisted electron conductions. {\it Thus,  the induction of topologically non-trivial
band structures does not immediately imply the topological properties of the non-equilibrium systems.}

On the other hand, the regime exists in which topological band structures can manifest themselves; 
we explicitly demonstrate that for {\it off-resonant} light where electrons cannot directly absorb photons,
the transport properties of the non-equilibrium systems attached to the leads are well approximated by 
the transport properties of the system described by  
the {\it static} effective Hamiltonian that incorporates the virtual photon absorption processes. 
In particular, the occupations of the electrons under this situation are close to the filling of the photon-dressed bands. 
As examples, we show that the transport properties under the application of off-resonant light is given by the photon-dressed Hamiltonian corresponding to a quantum Hall insulator without Landau levels\cite{Haldane1988} in the case of graphene, and to a gapped insulator with anomalous quantum Hall effects and magneto-electric response described by axion electrodynamics\cite{Qi2008} at the surfaces of three dimensional topological insulators. In these systems, the measurements in six-terminal configurations in \fref{figure1} lead to the near quantization of Hall conductance. Thus, the application of off-resonant circularly polarized light leads to an intriguing "Hall" effects {\it without applying a static magnetic field}. 

This paper is organized as follows. In \secref{sec:summary}, we describe the summary of the results, 
focusing on the analysis of graphene and three dimensional topological insulators 
under the application of off-resonant light. Here we provide the 
physical and intuitive explanations of the phenomenon of light-induced quantum Hall effects and 
refer to later sections for many important details. In \secref{sec:formalism}, we develop the formalism
for studying the non-equilibrium transport properties under periodical drives in many-terminal measurements. 
Our formalism is based on the extension of the multi-probe B\"uttiker-Landauer formula\cite{Buttiker1986}
to periodically driven systems, a "Floquet Landauer formula." 
We provide two distinct ways to calculate the transmission amplitudes in the driven systems. 
First method expresses the results in terms of the Floquet states and 
it illuminates the physical origin of the photon-assisted transport. 
Second method takes advantage of "Floquet Dyson's equation" to give an elegant solution which is more convenient
for numerical solution\cite{Martinez2003}. By taking the off-resonant limit of these solutions, the equivalence of 
transport properties under the application of light and those with effective photon-dresseed Hamiltonian is established. 

Most of the analysis in this paper assumes the absence of interactions among electrons as well as electron-phonon
interactions. We argue in \secref{sec:interaction} 
that, in the case of graphene and topological insulators under the off-resonant light, the 
results given in \secref{sec:summary} are robust against these interaction effects at low temperatures. 
While the measurements of transport properties require the attachment of leads, the probe of the effective gap
induced by light is plausible even in an isolated system. We propose in \secref{sec:pumpprobe} 
such measurements through the adiabatic
preparation of non-equilibrium systems combined with the transmission of probe laser with small frequencies. 
The essential ingredients in the arguments of \secref{sec:interaction} and \secref{sec:pumpprobe} are the 
extensions of adiabatic theorem and Fermi golden rule to periodically driven systems and Floquet states,
dubbed as "Floquet adiabatic theorem" and "Floquet Fermi golden rule." We give the detailed proof of 
these important statements in the Appendix. 
In \secref{sec:conclusion}, we conclude with possible extensions of this work.

\section{Summary of results} \label{sec:summary}

\subsection{Garphene effective Hamiltonian} 
Here we consider graphene as an example of semi-metals and study the change 
in the transport properties under the application of light. 
We model graphene by a hexagonal tight-binding model with two $\pi$-bands,
where we first neglect the electron-electron as well as electron-phonon interactions.
In \secref{sec:interaction} , we discuss the effects of these interactions 
and argue that they do not change the qualitative results of the analysis.
We consider the application of circularly polarized light perpendicular to the plane of 
graphene. For concreteness, here we represent the rotating electric field due to light 
as a time-dependent
vector potential $\vec{A}(t) =  A( \pm \sin(\Omega t), \cos(\Omega t) )$ with $\vec{E}(t) = \partial \vec{A}(t)/\partial t$,
where $\Omega$ is the frequency of light. Plus sign is for right circulation of light and minus sign for left circulation.  
The light intensity is characterized by the dimensionless number $\mathcal{A}= eAa/\hbar$ where $e$ is the electron charge 
and $a \approx 2.46 \AA$ is the lattice constant of graphene. 
For intensity of lasers and pulses available in the frequency regime of our interests $\sim 1000$THz, 
$\mathcal{A}$ is typically less than $1$. 
In this gauge, electrons accumulate phases as they hop in the lattice; 
\begin{equation}
H(t) = - J \sum_{<ij>,s} e^{iA_{ij}(t)} c_{i,s}^{\dagger} c_{j,s}, \label{hamiltonian}
\end{equation}
where $A_{ij}(t) =  e/\hbar (\vec{r}_{j} - \vec{r}_{i}) \cdot  \vec{A}(t)$ with $\vec{r}_{i}$ being the coordinates of 
the lattice site $i$,  $J$ is the hopping amplitude of electrons, and $s = \uparrow, \downarrow$ are spins of electrons. 
For simplicity, we only consider the orbital effect
of electromagnetic fields on electrons, and disregard the small Zeeman effect. 
The inclusion of the Zeeman effect is straightforward. 
In this limit, spins trivially double the Hilbert space and thus we suppress the spin indices in the following. 


When the light frequency is off-resonant for any electron transitions, 
light does not directly excite electrons and instead, effectively modifies the electron band structures 
through virtual photon absorption processes. 
Such off-resonant condition is satisfied for the frequency $\Omega \gg J$ in our model with $\pi$-bands.
More general case of on-resonant light can be analyzed through the formalism developed in later section
\secref{sec:formalism}.
The influence of such off-resonant light is captured 
in the static effective Hamiltonian $H_{\textrm{eff}}$\cite{Kitagawa2010a} defined through the evolution operator $U$ of the system after one period $T = 2\pi/\Omega$ as 
\begin{equation}
H_{\textrm{eff}} =\frac{i}{T} \log \left(U\right) \label{defeff}
\end{equation}
where $U= \mathcal{T} \exp\left(-i \int^{T}_{0} H(t) dt \right)$ and $\mathcal{T}$ is the time-ordering 
operator. 
Intuitively, $H_{\textrm{eff}} $ describes the dynamics of the system on time scales much longer than $T$. 
In the limit of $ \mathcal{A}^2 \ll 1$, 
$H_{\textrm{eff}}$ is particularly simple near the Dirac points;
 \begin{eqnarray}
H_{\textrm{eff}} & \approx & H_{0} + \frac{[H_{-1}, H_{1}]}{\Omega}+ O(\mathcal{A}^{4})  \label{heff} \\
&\approx & v_{G} ( \sigma_{y} k_{x} - \sigma_{x}  k_{y} \tau_{z}) \pm \frac{v_{G}^2\mathcal{A}^2}{\Omega} \sigma_{z} \tau_{z}+ O(\mathcal{A}^{4})\nonumber \\ 
&&\quad \quad \quad \quad \quad \quad \quad \quad \quad \textrm {(for infinite system)},  \nonumber
\end{eqnarray}
where $H_{n}$ is the discrete Fourier component of Hamiltonian, 
{\it i.e.} $H_{n} = \frac{1}{T}\int^{T}_{0} H(t) e^{it\Omega n} dt$. 
In the second line, $v_{G} = 3J/2$ is the velocity of Dirac electrons, $k_{x}$ and $k_{y}$ are momenta measured from the Dirac points,
$\sigma_{i}$ and $\tau_{i}$ are Pauli matrices representing sublattice and valley degrees of freedom, respectively.
\begin{figure}[t]
\begin{center}
\includegraphics[width = 8.5cm]{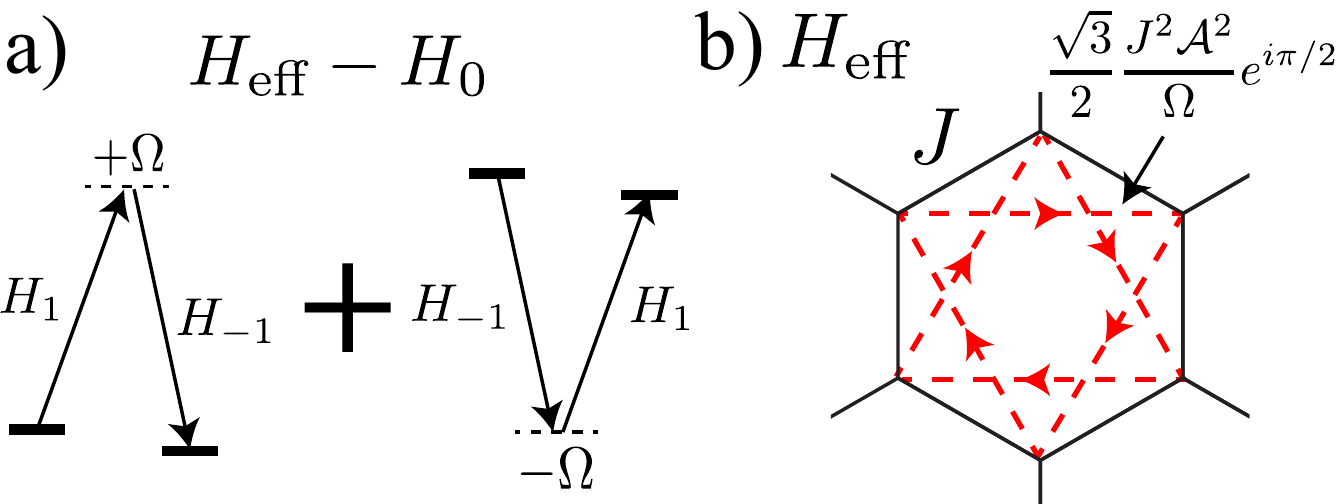}
\caption{ a) The modification of the Hamiltonian due to the virtual photon process can be intuitively understood 
as the sum of two second order processes where electrons absorbs and then emits a photon, and 
electrons first emit and absorb a photon. 
b) The illustration of the structure of $H_{\textrm{eff}}$ in real space for graphene under the application of 
right circularly polarized light in \eqnref{heff}. The commutator $[H_{1}, H_{1}]$ is the second neighbor hopping
 with phase $\phi =\pi/2$. Thus, the tight-binding model under the application of light realizes Haldane model 
 proposed in Ref. [\onlinecite{Haldane1988}]. }
\label{haldane}
\end{center}
\end{figure}
The modification of the Hamiltonian with respect to the static component $H_{0}$ is the second term in \eqnref{heff}.
This term can be easily understood as the sum of two second order processes as illustrated in \fref{haldane} a); 
one where electron absorbs a photon and then emits a photon $H_{1}\frac{1}{\omega -(\omega+\Omega)} H_{-1}$ where 
$\omega$ is the energy of the original electron;
another where electrons first emits a photon and then absorbs a photon, which leads to $H_{-1}\frac{1}{\omega- (\omega-\Omega)} H_{1}$. By summing these two contributions, we obtain the 
correction due to the second order process, given in the second term of  \eqnref{heff}.
In the second line, the plus sign is for right circulation of light polarization and the minus sign is for left circulation. 
For details of derivations, see \secref{sec:formalism}. We note that the expression of the effective Hamiltonian
in \eqnref{heff} is only valid in the gauge in which light is represented as time-dependent vector potential, 
and the effective Hamiltonian has different forms for other gauge such as the one in which 
light represented as time-dependent electric fields. 
The effect of virtual photon absorptions at the degenerate Dirac points is 
to open a gap with magnitude $\Delta= \frac{2 v_{s}^2 \mathcal{A}^2}{ \Omega}$. 
In \fref{figure2}, we illustrate the opening of the gap near one of the Dirac points upon the 
application of light for both infinite and finite systems.
This Hamiltonian $H_{\textrm{eff}}$ corresponds to 
a quantum Hall insulator, where each band is characterized by 
non-zero Chern number\cite{Haldane1988, Thouless1982}. 
 In \fref{figure2} b), we have plotted 
the spectrum of $H_{\textrm{eff}}$ where the system is infinite in $x$ direction and $150$ sites
in $y$ direction with armchair edges. 
Here, we have chosen the intensity and frequency of light to be $\mathcal{A}=0.3$ and $\Omega =7.5 J$. 
As a result of non-zero Chern number of the bands, 
$H_{\textrm{eff}}$ shows the existence of gapless chiral edge states, 
colored as blue and green, corresponding to the edge state in the upper and lower edge,
respectively.

It is instructive to write the effective Hamiltonian in \eqnref{heff} in real space. 
In the lowest order in $\mathcal{A}$, $H_{1}$ and $H_{-1}$ are 
the hopping between nearest neighbors with phase accumulations that depend on the 
direction of the hopping. Their commutators contain 
the second neighbor hopping with amplitudes $\frac{\sqrt{3}}{2} \frac{J^2 \mathcal{A}^2}{\Omega} e^{i\pi/2}$ 
as illustrated in \fref{haldane} b). Thus, the effective Hamiltonian is nothing but the Hamiltonian
proposed by Haldane\cite{Haldane1988} with sublattice potential $M=0$ and second hopping
strength $t_{2} = \frac{\sqrt{3}}{2} \frac{J^2 \mathcal{A}^2}{\Omega}$ with the flux $\phi=\pi/2$. 

Our results above differ from that of Inoue and Tanaka[\onlinecite{inoue2010}] in some important ways. 
In their work, the effect of circularly polarized light 
on the Haldane model\cite{Haldane1988} has been considered. They focused on the zero-photon 
sector of the Hamiltonian and concluded that the Chern number is zero whenever the second 
neighbor hopping $t_{2}$ is zero or the staggered magnetic flux $\phi$ is zero, as is presented in Eq. (7) of their paper. 
Their work showed no transition from topologically trivial band insulators to topological non-trivial bands with 
Chern numbers. Here we considered a simple tight-binding model, corresponding to $t_{2}=0$ and $\phi=0$
of Haldane model. In contrast with the result of Ref. [\onlinecite{inoue2010}], 
we show above (also see \secref{sec:effective}) that 
the virtual photon absorption and emission process represented by the second term of \eqnref{heff}
induces a gap at the Dirac point and leads to a non-zero Chern number. 
Such photon-dressing, which was neglected in the study of Ref. [\onlinecite{inoue2010}],
has dramatic effects at the degenerate Dirac points and should be taken into account.

There are a few different ways to probe the gap $\Delta$ in the effective Hamiltonian.
 For example, the gap opening might be confirmed through the observation of
the transmission of low frequency probe lasers. In \secref{sec:pumpprobe}, we propose the possibility
to probe this dynamically opened gap in an isolated system through adiabatic preparation of Floquet states. 
The main focus of this paper, however, is the study of the manifestations of the gapped Hamiltonian $H_{\textrm{eff}}$ in
many-terminal transport measurements depicted in \fref{figure1}, which we now describe. 


\begin{figure}[t]
\begin{center}
\includegraphics[width = 8.5cm]{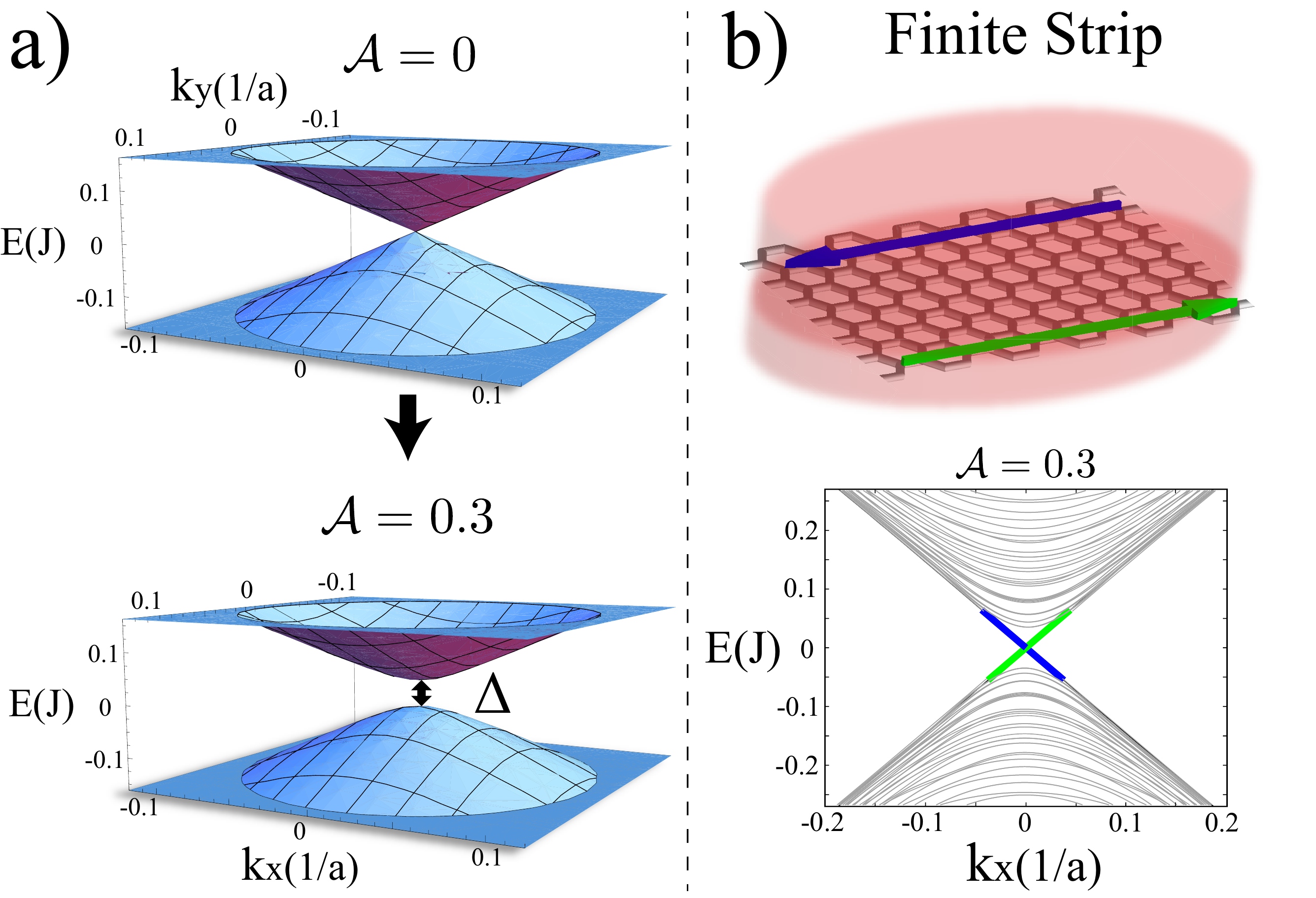}
\caption{ a) The spectrum of graphene for a single spin near one of the Dirac points for an infinite system. 
Here we plot the energy spectrum of a static system corresponding to $\mathcal{A} =0$(upper figure)
and the spectrum of $H_{\textrm{eff}}$ for the system under the application of light 
with light intensity $\mathcal{A} =0.3$(lower figure). A finite gap $\Delta$ opens at the Dirac point. 
b) The spectrum of $H_{\textrm{eff}}$ for a single spin near one of the Dirac points 
for infinite system in $x$ direction and $150$ sites in $y$ direction for $\mathcal{A}=0.3$. 
The frequency of light is chosen to be off-resonant with $\Omega = 7.5 J$. 
Here we chose the armchair edges, and 
in this case, the Dirac points are at $k_{x} =0$. 
$H_{\textrm{eff}}$ shows the existence of gapless chiral edge states for each spin originating from non-zero Chern numbers
of the bands, that are colored as blue and green, corresponding to the edge state in the upper and lower edge,
respectively. The propagation of chiral edge states for a single spin is illustrated in the upper figure.}
\label{figure2}
\end{center}
\end{figure}


\subsection{Floquet-Landauer formalism and Hall current}
\begin{figure}[t]
\begin{center}
\includegraphics[width = 8.5cm]{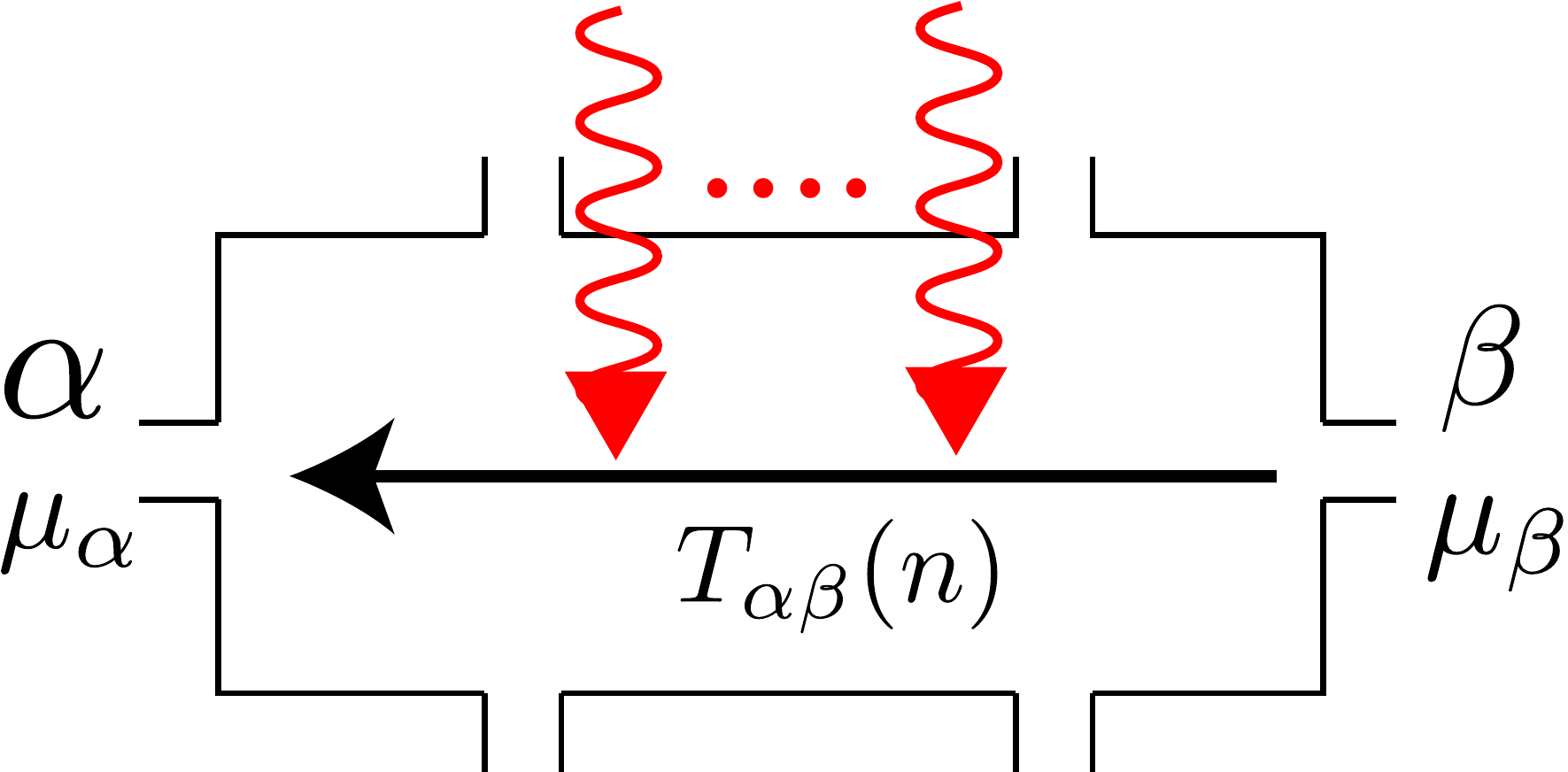}
\caption{ Illustration of the response conduction current in Floquet Landauer formula given in \eqnref{resdccurrent}. 
The transmission of the electrons can now happen with $n$ photon absorption/emissions. The total 
conductance is simply given by the sum of these contributions for each $n$.  }
\label{landauer}
\end{center}
\end{figure}

One of the central results of this paper is the demonstration
that measurements of DC current of the non-equilibrium system under the application of 
off-resonant light is determined by the static, photon-dressed Hamiltonian $H_{\textrm{eff}}$. 
To this end, we consider the many-terminal measurements of 
DC current under the application of light as in \fref{figure1}\cite{Buttiker1986}. 
The non-equilibrium transport properties of mesoscopic, periodically driven 
systems in this configuration have been studied previously, using Floquet theory\cite{sambe} combined with the Keldysh formalism
\cite{Jauho1994}. 
We express the general result obtained in these works as the extension of
the multi-probe B\"uttiker-Landauer formula\cite{Buttiker1986}
to periodically driven systems, a "Floquet Landauer formula"; 
\begin{eqnarray}
J^{\textrm{dc}}_{\alpha} &=& J^{\textrm{pump}}_{\alpha} + J^{\textrm{res}}_{\alpha} \label{dccurrent} \\
J^{\textrm{res}}_{\alpha} &=& \sum_{\beta} \left( \sum_{n } T_{\alpha \beta}(n) \right) (\mu_{\beta} - \mu_{\alpha}) 
 \label{resdccurrent} 
\end{eqnarray} 
Here $J^{\textrm{dc}}_{\alpha}$ is the DC component of the current at lead $\alpha$, 
$J^{\textrm{dc}}_{\alpha} = \frac{1}{T} \int^{T}_{0} J_{\alpha}(t) dt$. 
Here we assumed that the reservoirs are at zero temperature, and their chemical potentials $\{ \mu_{\alpha} \}$
are near the Dirac points, {\it i.e.} $\mu_{\alpha} \approx 0 $.
The DC current of a periodically driven system in \eqnref{dccurrent}
consists of two physically distinct contributions; the pump 
current\cite{Switkes1999}, $J^{\textrm{pump}}_{\alpha}$, which can be present even when 
all the reservoirs have the same chemical potential $\mu_{\alpha}$ and 
the response current, $J^{\textrm{res}}_{\alpha}$, which arises from the response of the driven system 
to the chemical potential differences of the reservoirs.
For the application of light we consider in this paper, 
the pump current $J^{\textrm{pump}}_{\alpha}$ is zero for inversion symmetric geometries, so we 
focus on the properties of the response current $J^{\textrm{res}}_{\alpha}$ in the following. 
The transmission coefficients $T_{\alpha \beta}(n)$ in \eqnref{resdccurrent} 
represent the transmission of electrons with energy $\mu_{\beta} \approx 0$ from lead $\beta$ to $\alpha$
during which electrons absorb (emit) $n$ photons as illustrated in \fref{landauer}. 
Thus, the response conduction in the presence of periodic drive can be seen as an extension of the static 
conduction, where the transmission can now happen with the absorptions/emissions of photons. 
We note that the expression of DC current in \eqnref{resdccurrent} is valid for arbitrary strength and frequency of the drives. 
The transmission probabilities $T_{\alpha \beta}(n)$ can be efficiently computed by dressing 
the propagators with photon absorptions/emissions\cite{Martinez2003}, and will be described in 
\secref{sec:formalism}. 
We emphasize that the response current 
is generally the sum of the contributions from $n$ photon absorption/emission processes 
(see \eqnref{resdccurrent}) and thus its 
transport property cannot be described by {\it any} static Hamiltonians.  The off-resonant case described 
below is an exceptionally simple case in this respect. 

We employ this Floquet-Landauer formalism to study 
the off-resonant, large frequency regime $J \ll \Omega$ with weak intensity of light, {\it i.e.} $ \mathcal{A}^2 \ll 1$.
In this regime, absorptions or emissions of photons are suppressed by $ \mathcal{A}^2$, and
 the transmission coefficients $T_{\alpha \beta}(n)$ with $n \neq 0$ is small and of the order of $ O( \mathcal{A}^2)$. 
 On the other hand, the zero-photon absorption/emission transmission coefficient $T_{\alpha \beta}(0)$ is modified due to virtual photon processes. Such modifications are included in $H_{\textrm{eff}}$ 
and the transmission probability is given by $T_{\alpha \beta}(0) = T^{\textrm{eff}}_{\alpha \beta}  + O( \mathcal{A}^2)$, 
 where $T^{\textrm{eff}}_{\alpha \beta}$ is the transmission probability of the static system described by $H_{\textrm{eff}}$. 
 These results will be rigorously established in \secref{sec:formalism}.
 
This correspondence demonstrates, under our assumptions, 
that graphene under the application of off-resonant light behaves as an insulator 
 with gap $\Delta= \frac{2 v_{s}^2 \mathcal{A}^2}{ \Omega}$ 
 with Hall conductance quantized at 
$2 e^2/h$ with possible corrections up to the order of $ O( \mathcal{A}^2)$.
Here the factor of $2$ comes from spin degrees of freedom. 
 While we established the results in the 
 perturbation theory on $\mathcal{A}$, it is possible to analytically 
 confirm the insulating behavior for all orders in $\mathcal{A}$ for weak contact couplings with leads(see \secref{sec:formalism}). 
We emphasize that although the effective Hamiltonian is perturbative in $\mathcal{A}$,
the Hall conductance at zero temperature is non-perturbative: an infinitesimal gap $\Delta$
is sufficient to yield a topological band with non-zero Chern number. 


A distinct feature of this light-induced Hall effect above is that the Hall conductance switches its
sign under the change of circulations of light polarization. 
This can be easily checked for the geometry of the system which is symmetric under $x \rightarrow -x$,
under which the circulation of light reverses. 
Such reversal of Hall current can be used in the experiments to 
distinguish this light-polarization {\it dependent} current from light-polarization {\it independent} current, which could 
originate from mechanisms we did not consider in this paper.

We briefly describe the requirements to observe the proposed phenomena with off-resonant light in graphene. 
The band width of graphene in the $\pi$ orbital is given by $6J$ where $J \approx 2.4$eV, placing the 
required frequency of off-resonant light to be soft x-ray regime with $\Omega = 3500 $THz. 
For this frequency of light, the gap of the system $\Delta$ can reach
$\Delta \approx 300$K for the strong light intensity $I \approx 3 \times 10^{12}$W/cm$^2$\cite{Schoenlein2000} which gives 
$\mathcal{A} \approx 0.09$, where we expect the Hall conductance to be quantized 
with possible correction of $1$\% of $2 e^2/h$.
In reality, even such high frequency of light is expected to be absorbed in graphene. 
Such direct electron excitations lead to reconfiguration of electron occupation numbers, which modifies 
the Hall conductance from its quantized values.

\subsection{Three dimensional topological insulators}
\begin{figure}[t]
\begin{center}
\includegraphics[width = 8.5 cm]{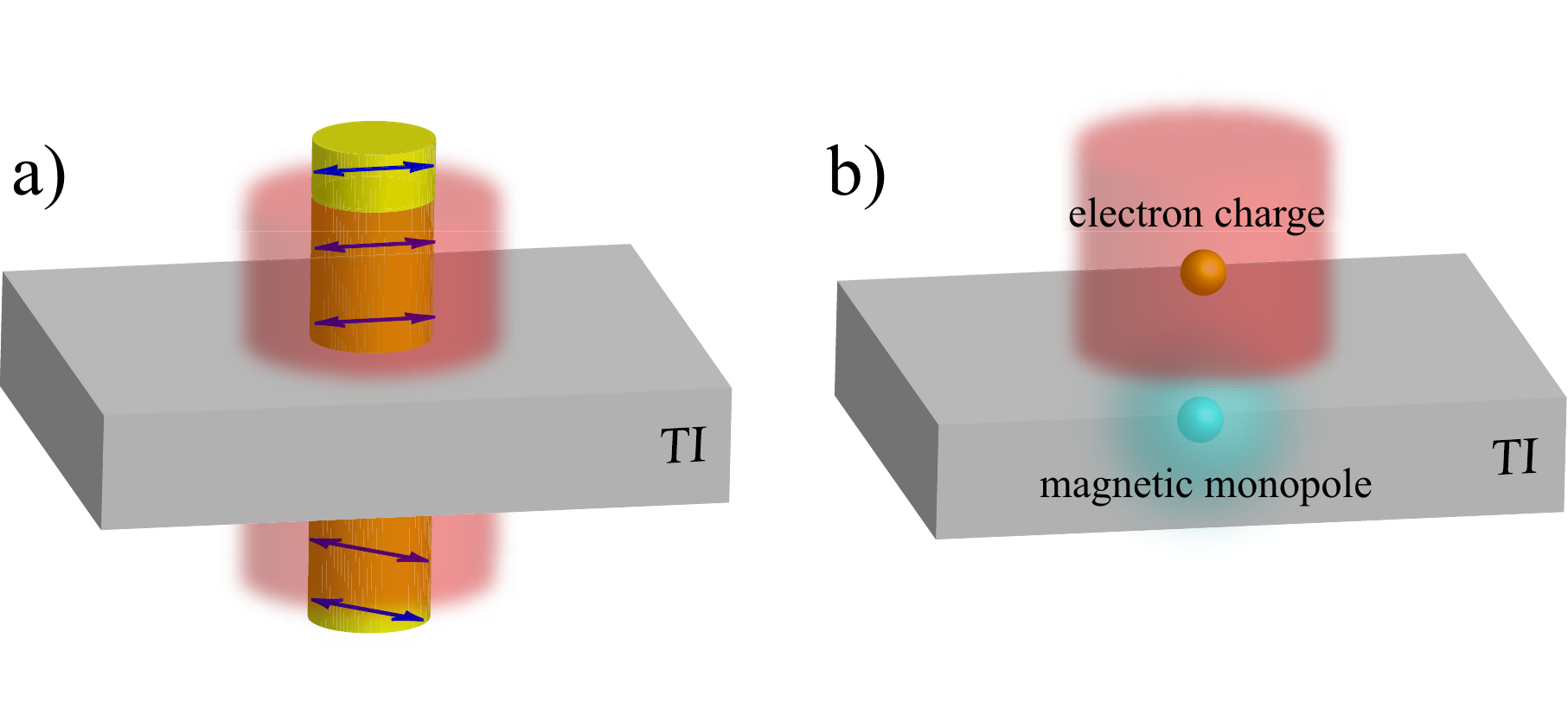}
\caption{ a) Measurements of Faraday/Kerr rotations in 
three dimensional topological insulator. Circularly polarized light with a large frequency is used to open a gap at the Dirac point
and linearly polarized light with small frequency within the gap is used to measure the Faraday/Kerr rotations. The polarization angle
of the light is denoted by blue arrows in the figure. 
b) The induction of magnetic monopole inside a three dimensional topological insulator through 
the application of light. 
Electron charge is placed near the surface, and circularly polarized light is applied on the surface to 
break the time-reversal symmetry and open the effective gap. The magnetic monopole is induced as a 
mirror image of the electron charge.   }
\label{figure3}
\end{center}
\end{figure}

The analysis of graphene above can be
directly extended to three dimensional topological insulators such as $\textrm{Bi}_{2}\textrm{Se}_{3}$. 
The low energy description of electrons on surfaces of 
~$\textrm{Bi}_{2}\textrm{Se}_{3}$ is given by two dimensional Dirac fermions\cite{Zhang2009}, 
and is described by the Hamiltonian $H^{\textrm{surf}} = v_{\textrm{TI}} (k_{x} \sigma_{y} - k_{y} \sigma_{x} )$
where $v_{\textrm{TI}}$ is the velocity of the Dirac fermion, and $\sigma_{i}$ are Pauli matrices 
corresponding to two bands near the Dirac point. 
As before, we assume the application of weak, off-resonant, circularly polarized light.
The orbital effect of the light is taken into account through the replacement 
$\vec{k} \rightarrow \vec{k} - \vec{A}(t)$. At the Dirac cone, the virtual photon process 
again opens a gap and the effective Hamiltonian is (see \eqnref{heff})
\begin{eqnarray}
H_{\textrm{eff}}^{\textrm{surf}} = v_{\textrm{TI}} (k_{x} \sigma_{y} - k_{y} \sigma_{x} ) \pm \frac{\mathcal{A}^2 v_{\textrm{TI}}^2}{\Omega} \sigma_{z} \label{heffti}
\end{eqnarray}
where $+(-)$ corresponds to the gap due to right(left) circularly polarized light. 
The consequences of the gap coming from the third term in \eqnref{heffti} are extensively 
investigated in Ref [\onlinecite{Fu2007, Qi2008}]. 
Just as in the case of graphene, the induced insulator is topologically non-trivial, and expected to 
result in anomalous quantum Hall effect with Hall conductance $ \pm \frac{e^2}{2h}$ with possible corrections up to the order of $ O( \mathcal{A}^2)$. 
Here we propose to probe the unique magneto-electric response of the gapped topological insulator through pump-probe
type measurements, where circularly polarized light is used to open a gap at the Dirac point
and linearly polarized light with small frequency within the gap is used to probe
the Faraday/Kerr rotations\cite{Okafaraday}, as illustrated in \fref{figure3} a) (also see \secref{sec:pumpprobe}). 
Unlike other schemes proposed previously with ferromagnetic layers, 
here the Faraday/Kerr rotations can only result from the topological insulators and they give unambiguous signature of 
magneto-electric effects. 
In a similar fashion, the existence of magnetic monopoles can be probed by placing an electric charge near the 
surface of the topological insulator in the presence of circularly polarized light (see \fref{figure3} b)). 

\subsection{Discussion}
In the analysis of graphene and topological insulators above, we assumed the off-resonance of light for entire bands, 
but the gap in the effective Hamiltonian opens whenever the light is off-resonant 
near the Dirac points, which requires much less stringent condition on the light frequency. 
However when inter-band electronic transitions occur due to photon
absorptions, subsequent relaxation processes are expected to change the electron occupation numbers
in the steady state, and modify  Hall conductance away from quantized values. 
Thus in the case of on-resonant light, the system is expected to display non-quantized Hall effects without magnetic fields. 
Moreover, as we show in \secref{sec:formalism}, the application of on-resonant light leads to the 
photo-assisted conductance and the resulting non-equilibrium transport property can no longer simply be 
described by the static effective Hamiltonian. The transport under the on-resonant light contains 
rich physics in itself, and it will be studied in the future works. 
On the other hand, it is possible to achieve the off-resonance with small frequency of light in, for example, 
the gapped systems such as Boron-Nitride by applying the sub-gap frequency of light. 
Many of these possible extensions can be studied through the formalism developed in the following sections.
The understanding and formalism obtained in this paper is likely to guide future
searches for the optimal systems to study photo-driven quantum
Hall effects without magnetic field.

\section{Non-equilibrium transport: formalism} \label{sec:formalism}
\subsection{ Floquet Landauer formula}
In this paper, we study the transport properties of systems under the application light 
in the Landauer-type configuration, where the systems are attached to the leads as in \fref{figure1}. 
Previous works\cite{Jauho1994} obtained the DC current in periodically driven systems 
in terms of the Floquet Green's functions $\hat{\mathcal{G}}(\omega, n)$ which represent the Fourier 
transform of retarded Green's function, and is a propagator
with frequency $\omega$ which absorbs (emits) $n$ photons.  
Here we express the general results in physically transparent fashion;
\begin{eqnarray}
J^{\textrm{dc}}_{\alpha} &=& J^{\textrm{pump}}_{\alpha} + J^{\textrm{res}}_{\alpha} \nonumber \\
J^{\textrm{pump}}_{\alpha} &=&  \sum_{n} \sum_{\beta} \int \frac{d\omega}{2\pi} T_{\alpha, \beta}(n, \omega)
 (f_{\alpha}(\omega) - f_{\alpha}(\omega + n\Omega)) \nonumber\\
J^{\textrm{res}}_{\alpha} &=&  \sum_{n} \sum_{\beta} \int \frac{d\omega}{2\pi}
T_{\alpha, \beta}(n, \omega)  (f_{\beta}(\omega) - f_{\alpha}(\omega)) \label{dcgreen}
\end{eqnarray}
\begin{equation}
  T_{\alpha, \beta}(n, \omega)  =   \Gamma_{\beta}(\omega) \Gamma_{\alpha}(\omega+n\Omega)   |\mathcal{G}_{j_{\alpha} j_{\beta}} (n, \omega)|^2 \label{transmission}
 \end{equation} 
 where $j_{\alpha}$ are the sites in graphene that are connected with leads $\alpha$, 
and $ \Gamma_{\alpha}(\omega) $ represents the coupling strength 
with leads $ \Gamma_{\alpha}(\omega)  = t_{\alpha}^2 \rho_{\alpha}(\omega)$ 
where $t_{\alpha}$ is the hopping strength from  graphene to lead $\alpha$ and $ \rho_{\alpha}(\omega)$ 
is the density of states in the lead $\alpha$ at energy $\omega$. Also, $f_{\alpha}(\omega)$ is the Fermi function 
 at lead $\alpha$, $f_{\alpha}(\omega) = \frac{1}{e^{\beta_{\alpha} (\omega -\mu_{\alpha})} +1}$ 
 where $\beta_{\alpha} =1/k_{B} T_{\alpha}$ is the 
 inverse temperature and $\mu_{\alpha}$ is the chemical potential of the reservoir connected to lead $\alpha$. 
 If we take the zero temperature limit and assume that differences of chemical potentials at each lead are small, 
 the expression in \eqnref{dcgreen} is reduced to the simpler Floquet Landauer formula given in \eqnref{resdccurrent}. 
 
 The calculation of conductance given by $J^{\textrm{res}}_{\alpha}$ reduces to the calculation of 
 Floquet Green's function $\hat{\mathcal{G}}(\omega, n)$. 
 Here, $\hat{\mathcal{G}}_{l,l'}(\omega, n)$ is nothing but a Fourier transform of the retarded Green's function 
 $G^{R}_{l,l'}(t, t')$.  Starting from the usual definition of the retarded Green's function,
 \begin{equation}
  G^{R}_{l,l'}(t, t') = -i \theta(t-t')( \braket{ c_{l}(t) c^{\dagger}_{l'}(t')} +   \braket{ c^{\dagger}_{l'}(t') c_{l}(t)} )
\end{equation}
we take the Fourier transform to obtain
   \begin{equation}
 G^{R}_{l,l'}(t, \omega) = \int^{\infty}_{-\infty} dt'  G^{R}_{l,l'}(t, t') e^{i(\omega+i 0^{+})(t-t')}. 
 \end{equation}
  Because we are driving the system at the given frequency $\Omega$, this Green's function, as a
 function of $t$, should contain only the discrete Fourier components. Therefore, we can expand as
   \begin{equation}
 G^{R}_{l,l'}(t, \omega) = \sum_{k=-\infty}^{\infty}  \mathcal{G}_{l,l'} (n, \omega) e^{-in\Omega t}
 \end{equation}
 
 The equation of motion followed by $\hat{\mathcal{G}}(\omega, n)$ can be obtained by writing 
 out the equation of motion for $G^{R}_{l,l'}(t, t')$ and taking its Fourier transform. 
 Th resulting equation can be written in the most compact form in the matrix equation whose 
 elements correspond to different (discrete) frequency components, $n\Omega$. 
 Explicitly, the equation is given by 
\begin{equation} 
(\omega + \bf{\Omega} - \bf{H}  -i \bf{\Gamma} /2) \bf{g} = I  \label{greeneom}
\end{equation} 
where 
 \begin{widetext}
  \begin{eqnarray} 
\bf{H} &=& \left(
\begin{array}{ccccc}  \ddots & \vdots  &  \vdots  & \vdots   &   \\ \cdots & \hat{H}_{0} & \hat{H}_{1} && 
\\    &  \hat{H}_{-1} & \hat{H}_{0} & \hat{H}_{1} &
 \\  & & \hat{H}_{-1} & \hat{H}_{0} &  \cdots \\  & \vdots  &  \vdots  & \vdots   &  \ddots
\end{array} \right), 
\bf{\Omega} = \left(
\begin{array}{ccccc}  \ddots  & &   &  &   \\ &1 \Omega& && 
\\    &  & 0 \Omega & &
 \\  & & & -1 \Omega &   \\  & & &   & \ddots 
\end{array} \right) \label{matrixdefinition} \\ 
\bf{\Gamma} &=& \left(
\begin{array}{ccccc}  \ddots  & &   &  &   \\ & \hat{\Gamma}(\omega+ \Omega)& && 
\\    &  &  \hat{\Gamma}(\omega) & &
 \\  & & & \hat{\Gamma}(\omega-\Omega) &   \\  & & &   & \ddots 
\end{array} \right), 
 \bf{g} =  \left(
\begin{array}{ccccc}  \ddots & \vdots  &  \vdots  & \vdots   &   \\ \cdots &\mathcal{G}(0, \omega)  & \mathcal{G}(1, \omega)  && 
\\    & \mathcal{G}(-1, \omega)  &\mathcal{G}(0, \omega)&  \mathcal{G}(1, \omega) &
 \\  & & \mathcal{G}(-1, \omega) &  \mathcal{G}(0, \omega)&  \cdots \\  & \vdots  &  \vdots  & \vdots   &  \ddots
\end{array} \right) \nonumber
  \end{eqnarray} 
  \end{widetext}
  $\bf{H}$ is the matrix of Hamiltonian, whose elements are 
  the Fourier components of Hamiltonian given by $\hat{H}_{n} = \frac{1}{T}\int^{T}_{0} \hat{H}(t) e^{it\Omega n} dt$. 
  $\bf{\Omega}$ is the diagonal matrix whose element is just the discrete frequency of the component. 
   $\bf{\Gamma}$ is a diagonal matrix that represents coupling of the systems with leads. 
   Its element $ \hat{\Gamma}(\omega)$ has non-zero value only at site $j_{\alpha}$ that couples with leads, 
   and with value $ \Gamma_{j_{\alpha},j_{\alpha}} (\omega)=  \Gamma_{\alpha}(\omega)  = t_{\alpha}^2 \rho_{\alpha}(\omega)$. Finally, $\bf{g}$ is the matrix composed of the Floquet Green's functions. 

 The equation in \eqnref{greeneom} can be thought of as the extension of the equation of motion for 
 free electrons coupled with leads to periodically driven systems. The rest of this section is devoted to
 the solution of this equation, and to the explanations of its physical significance for the response current 
given by \eqnref{dcgreen}. 
 
 In the following, 
 we give two different solutions of \eqnref{greeneom}. In \secref{sec:floquetstates}, we solve the equation
 by expressing the Green's functions in terms of Floquet states, the "stationary states" of periodically driven systems
 after one period of time. This solution illustrates the physical origin of the transport given in \eqnref{dcgreen}. 
In \secref{sec:effective}, we derive the equivalence of non-equilibrium transport and 
 the transport given by the effective photon-dressed Hamiltonian $H_{\textrm{eff}}$ claimed in \secref{sec:summary}
 by taking the off-resonant and weak intensity limit. 
 In  \secref{sec:dyson}, we give another solution of \eqnref{greeneom}, which is valid 
 for a certain class of periodic drive including the application of circularly polarized light. 
 This solution is derived by writing "Floquet Dyson's equation," and has the advantage of being numerically efficient. 
 Again, by taking the off-resonant, weak intensity limit of this solution, we arrive the result reported in \secref{sec:summary}.

 \subsection{Floquet states and Floquet Green's functions} \label{sec:floquetstates}
 \subsubsection{General relation}
 
The "stationary states" of the Schr\"odinger equation for periodically driven systems are the states 
which return to themselves after one period of time, $T= 2\pi/\Omega$, with possible phase accumulations. 
These so-called Floquet states are the eigenstates of the evolution operator over one period,
and thus also eigenstates of effective Hamiltonian $H_{\textrm{eff}}$ defined in \eqnref{defeff}. 
Green's functions $\hat{\mathcal{G}}(\omega, n)$ 
that describes the propagation of particles with possible absorptions/emissions of 
photons have natural expressions in terms of these Floquet states. 

The time evolution of the Floquet states $\ket{\phi_{a}(t)}$
can be expanded in the discrete Fourier component of the driving frequency $\Omega$, and can be expressed as 
\begin{equation} 
\ket{\phi_{a}(t)} = e^{-iE_{a} t} \sum_{n} e^{-i\Omega n t} \ket{\phi^n_{a}} \label{physicalfloquet}
\end{equation}
where $E_{a}$ is the quasi-energy of the Floquet state $\ket{\phi_{a}}$ for $H_{\textrm{eff}}$
and $\ket{\phi^n_{a}}$ is $n$th Fourier component of the Floquet state.
The eigenstates of effective Hamiltonian $H_{\textrm{eff}}$ are given as 
$\ket{\phi_{a}} = \sum_{n} \ket{\phi^n_{a}}$. 
As one can see from the expression above, the quasi-energy $E_{a}$ is only well-defined up to
the driving frequency $\Omega$, {\it i.e} we can equally define $E_{a} + m\Omega$ as the quasi-energy
of $\ket{\phi_{a}(t)}$ by redefining $\ket{\phi^n_{a}} \rightarrow \ket{\phi^{n-m}_{a}}$. Physically, this means
the quasi-energy is only conserved up to the driving frequency $\Omega$ because the system can absorb or emit 
photon energies. Also, this fact can be seen as 
a natural consequence of the breaking of continuous time-translation invariance through external drivings where
the system only possesses discrete time-translation invariance under $t \rightarrow t+T$. 
In the following, we assume that $-\Omega/2  \leq E_{a} \leq \Omega/2$ without loss of generality. 
We take the normalization of 
Floquet states such that $ \sum_{n} \braket{\phi^n_{a}| \phi^{n}_{b}}  =   \delta_{a b}$.
The Schr\"odinger equation for the Fourier components of the Floquet states is time-independent, 
\begin{equation} 
(E_{a} + n\Omega) \ket{\phi^n_{a}} = \sum_{m} H_{n-m} \ket{\phi^m_{a}} \label{eom}
\end{equation}
where $H_{n}$ is the discrete Fourier component of Hamiltonian, 
{\it i.e.} $H_{n}= \frac{1}{T} \int^{T}_{0} H(t) e^{it\Omega n} dt$.
This equation encapsulates the evolution 
of states that allows the absorptions/emissions of photons; application of Hamiltonian $H_{m}$ leads to the 
absorption of $m$ photons and the state $\ket{\phi^{n}}$ is the component of the state with 
$n$ photons.  
Here we considered the evolution of the systems in the absence of coupling with leads, but 
in the presence of the coupling with leads, zero frequency component of the Hamiltonian $H_{0}$ 
contains the imaginary "leaking" term $i\Gamma/2$. 

This Schr\"odinger equation takes, in the matrix form, 
\begin{equation}
E_{a} \ket{\phi_{a}} =  (- \bf{\Omega} + \bf{H} ) \ket{\phi_{a}} \label{floqueteom}
\end{equation}
where 
\begin{equation}
 \ket{\phi_{a}}  =  \left(
\begin{array}{c} \vdots  \\  \ket{\phi^1_{a}} \\ \ket{\phi^0_{a}} \\ \ket{\phi^{-1}_{a}} \\ \vdots  
\end{array} \right)
\end{equation}
The Hamiltonian matrix  $\bf{H}$ and driving frequency matrix  $\bf{\Omega}$ are given in \eqnref{matrixdefinition}.

Thus, Floquet state $ \ket{\phi_{a}} $ is nothing but the eigenstates of the composite Hamiltonian
$(- \bf{\Omega} + \bf{H} ) $. Notice that "shifted" state 
\begin{equation}
 \ket{\phi_{a}^{\textrm{shift}}(n)}  =  \left(
\begin{array}{c} \vdots  \\  \ket{\phi^{-n+1}_{a}} \\ \ket{\phi^{-n}_{a}} \\ \ket{\phi^{-n-1}_{a}} \\ \vdots  
\end{array} \right) \label{nfloquet}
\end{equation}
is also an eigenstate with eigenvalue $E_{a} + n\Omega$. 

Now we can relate the Floquet eigenstates given by \eqnref{floqueteom} and Floquet Green's function
given by \eqnref{greeneom}. The formal solution of \eqnref{greeneom} is obtained in terms of 
the eigenstates $\ket{\phi_{a}}$ of the matrix $\bf{\Omega} - \bf{H}  -i \bf{\Gamma} /2$ as 
\begin{equation}
\bf{g} = \sum_{a} \frac{ \ket{\phi_{a}} \bra{\tilde{\phi}_{a}}} {\omega - \epsilon_{a}} \label{formals}
\end{equation}
Here, $\bra{\tilde{\phi}_{a}}$ is the state that is determined from 
$\braket{ \tilde{\phi}_{a} | \phi_{b}} = \delta_{ab}$. (Note that in the presence of 
$i \bf{\Gamma} /2$, $\bra{\tilde{\phi}_{a}}$ is not just a complex conjugate of $\ket{\phi_{a}}$.)
As is clear from \eqnref{floqueteom}, the eigenstates of 
$\ket{\phi_{a}}$ of the matrix $\bf{\Omega} - \bf{H}  -i \bf{\Gamma} /2$ is nothing but the 
Floquet states in the presence of the coupling with leads, represented by $-i \bf{\Gamma} /2$.
Notice that the eigenstates $\ket{\phi_{a}}$ in \eqnref{formals} include all the shifted states 
$ \ket{\phi_{a}^{\textrm{shift}}(n)} $ in \eqnref{nfloquet} for all integers $n$. With this understanding, we 
obtain the expression of the Floquet Green's function $\hat{\mathcal{G}}(n, \omega)$
as 
\begin{equation}
  \hat{\mathcal{G}}(n, \omega)= 
  \sum_{a} \sum_{m} \frac{ \ket{\phi^{n-m}_{a}} \bra{\tilde{\phi}^{-m}_{a}} } { \omega - E_{a} -m\Omega} 
  \label{green}
\end{equation}
We can see that Floquet Green's function is an intuitive extension of the Green's function 
for free electrons that allows absorptions and emissions of photons. As is expected,
this Green's function transfers the state in $-m$ photon sector $ \bra{\tilde{\phi}^{-m}_{a}}$ to 
$n-m$ photon sector $ \ket{\phi^{n-m}_{a}} $ by absorbing $n$ photons. 

In the presence of on-resonant light, Floquet states generally contain non-zero amplitudes in 
$ \ket{\phi^{n}_{a}}$ for more than one value of $n$, and therefore, the contributions to the 
response current in \eqnref{resdccurrent} from a few photon absorptions/emissions are non-zero. 
Thus effective static Hamiltonian or band structures, which are only the description of average of $n$ photon
states $\sum_{n} \ket{\phi_{a}^n}$, does not appropriately capture 
transport properties under the on-resonant light. In this case, it is necessary to compute the full Floquet 
Green's function given by \eqnref{green} and calculate the response current in \eqnref{dcgreen}. 

On the other hand, in the case of off-resonant, weak intensity of light, transport properties of non-equilibrium
systems can be described by an effective photon-dressed Hamiltonian. In the next section, we provide the proof 
in the case of semi-metals such as graphene and topological insulators. 

\subsubsection{Effective Hamiltonian description} \label{sec:effective}
As summarized in \secref{sec:summary}, a rich physics appears when circularly polarized light is 
applied to graphene and topological insulators. 
The description of the non-equilibrium transport takes a particularly simple form 
for the off-resonant light in the limit of small light intensity $\mathcal{A} \ll 1$.
Here we apply the general formalism developed in the previous section to these systems
and study the transport property by obtaining the Floquet Green's function 
in \eqnref{green}. 

From the explicit form of the Hamiltonian in \eqnref{hamiltonian}, it is clear that $H_{n}\sim O(\mathcal{A}^{|n|})$. 
This simply means the absorptions of photons are suppressed by the factor $O(\mathcal{A}^{|n|})$.
Thus, the $n$ photon sectors of the Floquet states $\ket{\phi^n_{a}}$ are expected to scale as 
$ \ket{\phi^n_{a}} \sim O(\mathcal{A}^{|n|})$ with 
 zeroth order solution being the static part of the Hamiltonian $H_{0} + i\Gamma/2$. 
Here the term $i\Gamma/2$ represents the coupling with leads
and we assumed the same strength of the coupling $\Gamma= \Gamma_{\alpha}$ at each lead $\alpha$
and further assumed that it is independent of frequency. This latter assumption is not important 
in the off-resonant case because current is essentially conducted only at chemical
potential of leads, as we will confirm later. 
 Starting from the equation \eqnref{eom}, we apply a degenerate perturbation theory 
 in the lowest non-trivial order in $\mathcal{A}$ to obtain 
 \begin{eqnarray}
\left( H_{0} + i\Gamma/2 + \frac{[H_{-1}, H_{1}]}{\Omega} \right) \ket{\phi^0_{a}} &=& E_{a}  \ket{\phi^0_{a}}  \\
 \ket{\phi^n_{a}} &=& \frac{1}{n\Omega} H_{n} \ket{\phi^0_{a}} \quad  \textrm{for }n \neq 0 \nonumber \\ \label{floquetstate}
 \end{eqnarray}
 In the derivation, we assumed $E_{\alpha} \ll \Omega$, so the expression above is only valid near 
 the Dirac points. Note that since the Hamiltonian $H_{0}$ is degenerate at the Dirac points, 
the mixings of the states due to the perturbations of $\mathcal{A}$ are not small.
 This result indeed shows that $\ket{\phi^n_{a}} \sim O(\mathcal{A}^{|n|})$, 
 and therefore, the Floquet states $\ket{\phi_{a}}$ can be approximated by 
 the zeroth level of the Floquet states $\ket{\phi^0_{a}}$, 
 which is given by the eigenstates of the effective Hamiltonian
 $H_{\textrm{eff}} =H_{0} + \frac{[H_{-1}, H_{1}]}{\Omega} $ plus the coupling with leads $i\Gamma/2$. 
 
 Using the solution of Floquet states above, we can obtain the response current in the lowest order in 
 $\mathcal{A}$. 
The scaling $\ket{\phi^{n}_{a}} \sim O(\mathcal{A}^n)$ in \eqnref{floquetstate} directly implies that 
 $| \hat{\mathcal{G}}(n, \omega)|^2 \sim O(\mathcal{A}^{2n})$. Moreover, 
Green's function with no photon absorptions or emissions can be approximated as 
\begin{eqnarray}
  \hat{\mathcal{G}}(0, \omega) &=&  
  \sum_{a}  \frac{ \ket{\phi^{0}_{a}} \bra{\tilde{\phi}^{0}_{a}} } { \omega - E_{a} } + O(\mathcal{A}^2) \nonumber  \\
 & \equiv&   \hat{\mathcal{G}}^{\textrm{eff}}(\omega) + O(\mathcal{A}^2) \nonumber
\end{eqnarray}
where $\ket{\phi^{0}_{a}} $ is the eigenstates of  $H_{\textrm{eff}}+ i\hat{\Gamma}/2$
, and therefore $\hat{\mathcal{G}}^{\textrm{eff}}(\omega)$ is the 
free electron Green's function for the static system with Hamiltonian $H_{\textrm{eff}}$ coupled
with leads. Thus, these arguments combined with the expressions of currents in \eqnref{dcgreen} and \eqnref{transmission}
prove that the many-terminal measurements of non-equilibrium systems under the off-resonant light 
give the same result as if the system is given by the static effective Hamiltonian $H_{\textrm{eff}}$. 
We emphasize that this result immediately implies the following two facts; 
1. the non-equilibrium system displays insulating behaviors in the longitudinal conductance
2. Hall conductance is nearly quantized with possible correction of $ O(\mathcal{A}^2) $, as 
presented in \secref{sec:summary}. 

\subsubsection{Insulating behavior for gapped effective Hamiltonian $H_{\textrm{eff}}$} \label{sec:insulator}
In the analysis of the previous section, 
we established the insulating behaviors of graphene under the application of off-resonant 
light through the perturbation theory in $\mathcal{A}$. Such analysis only shows that the longitudinal conductivity 
is small and of the order of $\mathcal{A}^2$, but does not show, in a strict sense, that the conductivity goes to zero 
at zero temperature. Using the formalism developed in previous sections, it is possible to show that the 
non-equilibrium system is an insulator for all orders in $\mathcal{A}$ as long as the chemical potential
of leads lies below the effective gap of $H_{\textrm{eff}}$. The argument does not rely on the off-resonant 
condition and in principle applicable whenever $H_{\textrm{eff}}$ has a gap. 

\begin{figure}[t]
\begin{center}
\includegraphics[width = 7cm]{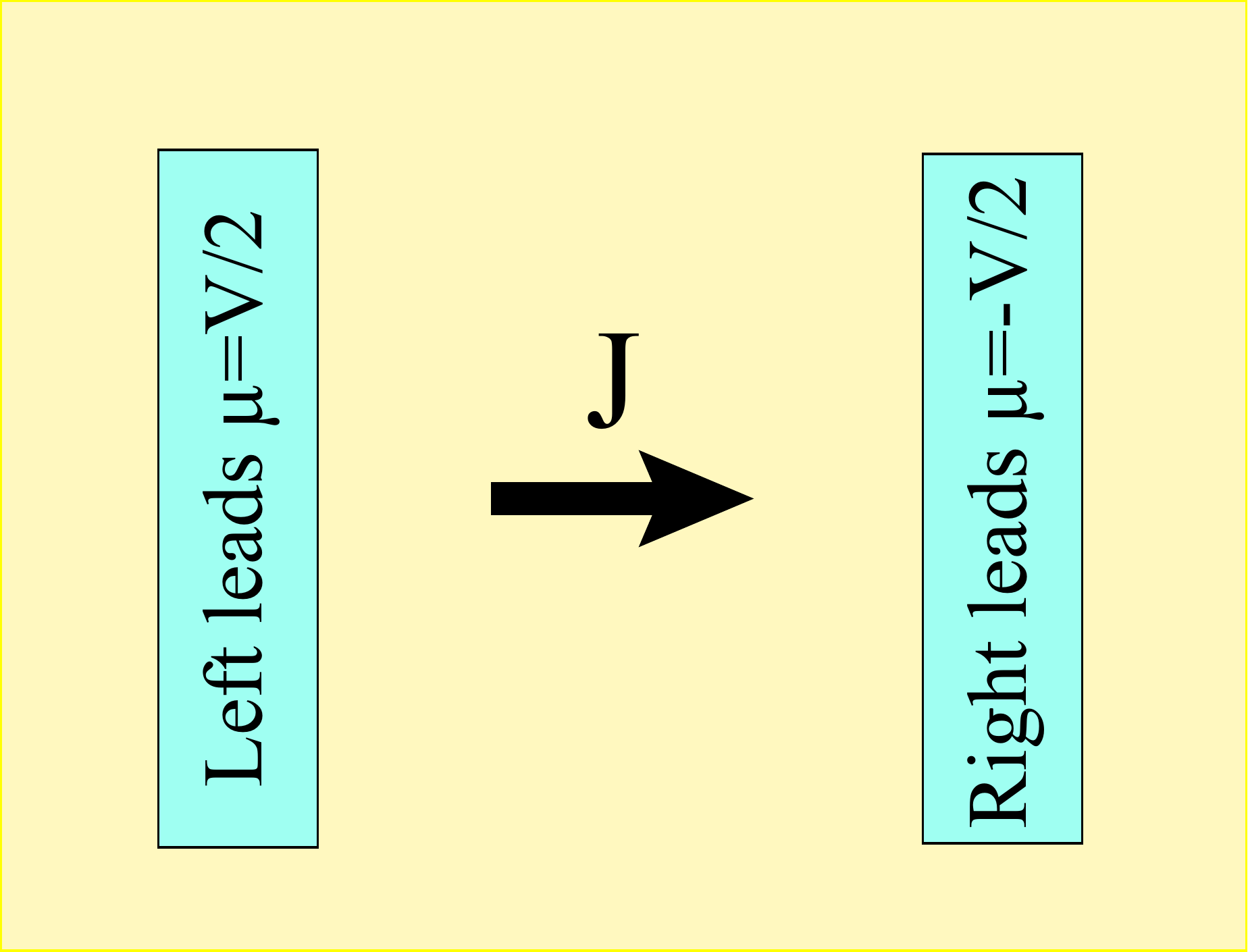}
\caption{ Illustration of the configuration considered in \secref{sec:insulator} to demonstrate 
the insulating behavior for gapped effective Hamiltonian $H_{\textrm{eff}}$.
$N_{L}$ number of leads are attached to the infinite plane of graphene as "left" leads and 
$N_{R}$ number of leads are attached as "right" leads. 
The chemical potentials of the reservoirs connected to left leads are assumed to be at $V/2$
and those of the reservoirs connected to the right leads are at $-V/2$ with $|V| \ll J$ }
\label{figure4}
\end{center}
\end{figure}

Here we consider an infinite plane of graphene, and we attach $N_{L}$ number of leads as "left" leads and 
$N_{R}$ number of leads as "right" leads, where these leads are separated by a large distance, see Fig.~\ref{figure4}. 
Here we assume that the leads are coupled with the system with equal strength, given by $\Gamma(\omega)$. 
For clarity, we consider the situation in which the chemical potentials of the reservoirs connected to left leads are at $V/2$
and those of the reservoirs connected to the right leads are at $-V/2$ with $|V| \ll J$. 

In the limit of small coupling strength, the Green's function in \eqnref{green} can be obtained
through the perturbation theory on $\Gamma$, and are given by 
\begin{equation}
  \hat{\mathcal{G}}(n, \omega) \approx 
  \sum_{a} \sum_{m} \frac{ \ket{\phi^{n-m}_{a}} \bra{\phi^{-m}_{a}} } { \omega - E_{a} -m\Omega- 
 i  \gamma_{a}( \omega) } 
\end{equation}
  where $\gamma_{a}(\omega) = \sum_{n} \braket{\phi_{a}^{n}|
   \hat{\Gamma}(\omega+n\Omega)/2|\phi_{a}^{n}}$, and $ \ket{\phi^{n}_{a}}$ and $E_{a}$ 
   are the Floquet states and (quasi-)energy of the system in the absence of the coupling with leads. 
  In the limit of small $\gamma_{a}(\omega)$, the square of the Green's function $ \hat{\mathcal{G}}(n, \omega)$ can be 
  approximated by a delta function, so that the transmission probability also becomes a delta function in frequency; 
\begin{eqnarray*}
T_{L, R}(n, \omega) &=&
 \sum_{L_{i}, R_{i}} \sum_{a, m} \Gamma(\omega) \Gamma(\omega+n\Omega) \\
 && \times
  \frac{ |\braket{ j_{L_{i}}| \phi^{n-m}_{a} }|^2 |\braket{ \phi^{-m}_{a} | j_{R_{i}} }|^2 }{ 2 \gamma_{a}(\omega)}
\pi  \delta(\omega- E_{a} - m\Omega)  \label{longitudinal}
\end{eqnarray*}
where $j_{L_{i}(R_{i}) }$ are the sites of left (right) leads. 
Now note that the (quasi-)energies $E_{a}$ are the eigenvalues of $H_{\textrm{eff}}$ in Eq. (2) in the main text. 
Therefore, if all the chemical potentials lies within the gap of $H_{\textrm{eff}}$, {\it i.e.} $|V| \leq \Delta$ 
the delta function gives 
zero everywhere for $-V/2 \le \omega \le V/2$. 
Note that we have taken the quasi-energies $E_{a}$ to lie between $-\Omega/2 \leq E_{a} \Omega/2$ and thus, 
by assumption, $\Delta \leq \Omega$. $m\Omega$ term in the delta function of \eqnref{longitudinal} 
accounts for the possible transmission of electrons at high/low energies through photon absorption/emission 
processes. As long as we are interested in the transmission of electrons near the chemical potential which 
lies within the effective gap, $m\Omega$ term plays no role in the conduction. 
Thus, from the expression of the DC current in \eqnref{dcgreen}, it is clear that the current has to be 
zero when $|V| \leq \Delta$. 

The argument above is general and did not require the condition of off-resonance. In the case of 
on-resonant light, the effective band structures are given by mixing the static eigenstates whose energies 
differ by $\Omega$. This "folding" of the band structures generically leads to a large number of states 
appearing in the effective Hamiltonian near the chemical potential, and subsequently the effective gap in 
$H_{\textrm{eff}}$ becomes proportional to $O(\mathcal{A}^n)$ with $n$ approximately determined by 
the ratio of the static band width and the driving frequency $\Omega$. 
While the insulating behavior should be observable in the small window
of the gap, the gap could be small in this case. 

\subsection{Floquet Dyson's equation} \label{sec:dyson} 
In this section, we present yet another way to obtain the Floquet Green's function $\hat{\mathcal{G}}(n, \omega)$
which gives an efficient way to numerically evaluate the Floquet Green's function for a certain class of 
periodical drives. 

In this section, we consider the Hamiltonian that depends only on the 
first harmonics of the driving frequency $\Omega$, namely, the Hamiltonian takes the form
\begin{eqnarray}
H(t) = H_{0} + V_{1} e^{-i\Omega t} + V_{-1} e^{i\Omega t}. \label{continuedhamiltonian}
\end{eqnarray}
For example, for the application of the circularly polarized light to two dimensional lattice systems, 
$V_{1} = \sum_{j} (x_{j} + iy_{j}) c_{j}^{\dagger} c_{j} $ and $V_{-1} =V_{1}^{\dagger}$ in the gauge 
in which the light is represented as a circulating potential. However, in this gauge,
$V_{\pm 1}$ diverges as $x_{j}, y_{j} \rightarrow \infty$, care must be taken to study with this gauge. 
Conceptually useful gauge is the gauge in which
the effect of light is represented as a phase accumulation as in \secref{sec:summary}. For weak amplitude of light, 
we can approximate the Hamiltonian in this gauge in the form of \eqnref{continuedhamiltonian} with $V_{1} = H_{1}$
and $V_{1} = H_{-1}$. 
As before, we are interested in the terminal measurements of conductance, and thus 
assume that the static part of the Hamiltonian $H_{0}$ contains 
the "leaking" of particles into leads given by $i \Gamma(\omega)/2$. 

\begin{figure}[t]
\begin{center}
\includegraphics[width = 8.5 cm]{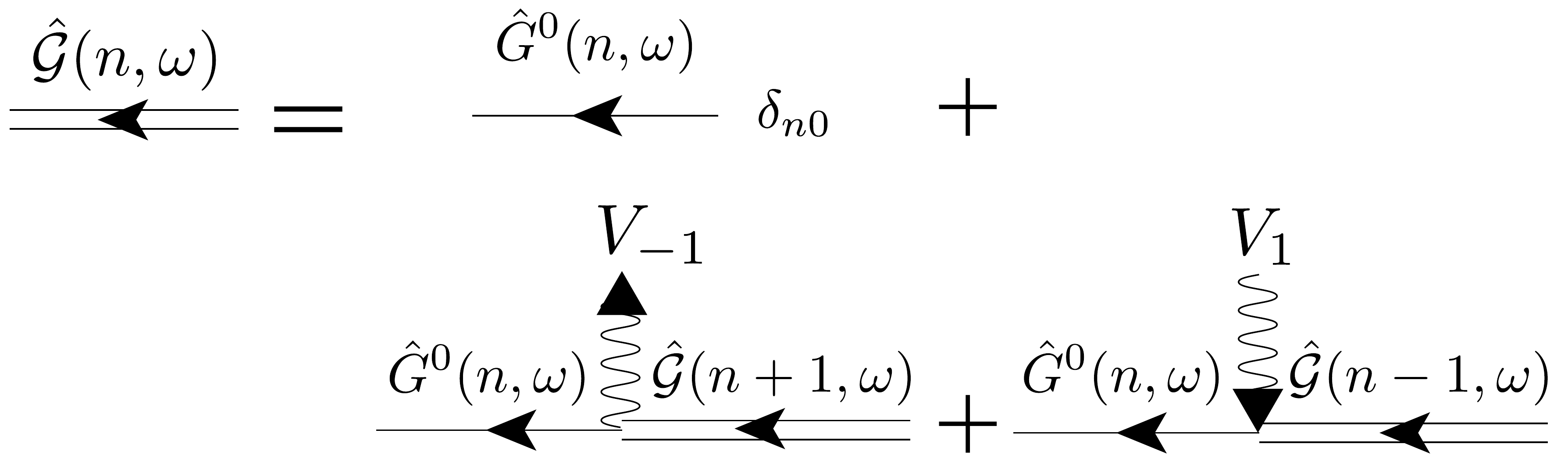}
\caption{The Floquet Dyson equation. The propagator goes from right to left. Double line represents the 
full propagator $\hat{\mathcal{G}} (n, \omega)$ and single line is a bare propagator 
$\hat{\mathcal{G}} (n, \omega)= \frac{1}{ \omega + n\Omega -H_{0}}$ which does not include the 
effect of photon absorptions. }
\label{dyson}
\end{center}
\end{figure}

In order evaluate Floquet Green's functions, we first rewrite the equation
\eqnref{greeneom} in the suggestive form of "Floquet Dyson's equation" (see \fref{dyson}); 
  \begin{eqnarray}
  \hat{\mathcal{G}} (n, \omega) &=& \delta_{n0}  \hat{G}^{0}(n,\Omega)+  \nonumber \\
   && \hat{G}^{0}(n,\Omega) \left( V_{-1} \hat{\mathcal{G}}(n+1, \omega) + V_{1} \hat{\mathcal{G}}(n-1, \omega)  \right) \nonumber \\
 \end{eqnarray}
 Here $\hat{G}^{0}(n,\Omega)= \frac{1}{ \omega + n\Omega -H_{0}}$ represents the bare propagator of 
 a particle with $n$ photons. 
 This equation has the intuitive understanding of the full propagator $ \hat{\mathcal{G}} (n, \omega) $ that 
 represents the $n$ photon absorption process as being composed of the full propagation of 
 $ \hat{\mathcal{G}} (n \pm 1, \omega) $ followed by the absorption or emission of a photon, followed by 
 the propagation of the bare particle. 
 
 A particularly elegant solution for $  \hat{\mathcal{G}} (n, \omega)$ is provided by continued fraction 
 method\cite{Martinez2003}. The building block of the solution is the dressed propagator 
 \begin{eqnarray}
   \hat{\mathcal{F}}_{+} (n, \omega) = \frac{1}{ \left(G^{0}\right)^{-1}(n, \omega) 
   - V_{-1} \frac{1}{\left(G^{0}\right)^{-1}(n+1, \omega) - V_{-1} \frac{1}{\cdots} V_{1}} V_{1} } \nonumber \\
   \textrm{for $n>0$} \nonumber \\
      \hat{\mathcal{F}}_{-} (n, \omega) = \frac{1}{ \left(G^{0}\right)^{-1}(n, \omega) 
   - V_{1} \frac{1}{\left(G^{0}\right)^{-1}(n-1, \omega) - V_{1} \frac{1}{\cdots} V_{-1}} V_{-1} } \nonumber \\
   \textrm{for $n<0$} \nonumber 
   \end{eqnarray}
 The propagator    $\hat{\mathcal{F}}_{+} (n, \omega) $ is dressed only from the higher photon number
 states, and  the propagator   $\hat{\mathcal{F}}_{-} (n, \omega) $ is dressed by the lower photon number states. 
 The full propagator is then given as 
 \begin{eqnarray}
 \hat{\mathcal{G}} (0, \omega) &=& \left( \omega - H_{0} - V_{\textrm{eff}} \right)^{-1} \\
 V_{\textrm{eff}} &=&  V_{1}  \hat{\mathcal{F}}_{-} (-1, \omega) V_{-1} +  V_{-1}  \hat{\mathcal{F}}_{+} (1, \omega) V_{1} \nonumber \\
 \hat{\mathcal{G}} (n, \omega) & =& \hat{\mathcal{F}}_{+} (n, \omega) V_{1} \cdots \hat{\mathcal{F}}_{+} (1, \omega) V_{1}
  \hat{\mathcal{G}} (0, \omega)   \nonumber  \\
  &&  \quad \quad \quad  \quad \quad \quad   \quad \quad \quad  \quad \quad \quad \textrm{for $n>0$}\nonumber  \\
 & =& \hat{\mathcal{F}}_{-} (n, \omega) V_{-1} \cdots \hat{\mathcal{F}}_{-} (-1, \omega) V_{-1}
  \hat{\mathcal{G}} (0, \omega) \nonumber \\
 &&  \quad \quad \quad  \quad \quad \quad   \quad \quad \quad  \quad \quad \quad \textrm{for $n<0$}\nonumber  \\
  \end{eqnarray}
  This solution is valid for any driving frequency. 
  Remarkably, we see that the zero-photon absorption propagator $ \hat{\mathcal{G}} (0, \omega)$ is simply 
  given by the propagator in an effective Hamiltonian $H_{\textrm{eff}} = H_{0}+ V_{\textrm{eff}}$. 
  
  For the gauge in which light is represented as time-dependent vector potential, and weak intensity of light
  $\mathcal{A} \ll 1$, we can approximate 
  $ \hat{\mathcal{F}}_{+} (1, \omega) =  \hat{\mathcal{F}}_{-} (-1, \omega) = \frac{1}{\Omega}$ in the limit of high frequency. Thus, we reproduce the result we obtained in \secref{sec:effective} of the effective Hamiltonian 
$H_{\textrm{eff}} =H_{0} + \frac{[H_{-1}, H_{1}]}{\Omega}$ in this limit.

\section{Effect of electron-electron and electron-phonon interactions} \label{sec:interaction}
The non-equilibrium transport properties described in \secref{sec:summary} are robust against 
interactions such as electron-electron interactions, interactions between electrons and disorder,
and electron-phonon interactions. 
The electron-electron interactions only renormalize 
the velocity of Dirac electrons, $v_{G}$ and $v_{\textrm{IT}}$, and do not change 
the Dirac nature of the electrons near the Fermi surface\cite{Novoselov2005}.
The quantum Hall insulators are insensitive to disorders due to the topological origin of the phase, as long as the disorder strength is small compared to the gap size, $\Delta$\cite{Thouless1982}. 

The robustness of the phenomena against phonon scatterings originates from the conservation of 
energy in $H_{\textrm{eff}}$ up to the light frequency $\Omega$. 
When the chemical potentials of leads lie in the gap of $H_{\textrm{eff}}$, 
the non-equilibrium current in many-terminal measurements is conducted through electrons in the lower band
of $H_{\textrm{eff}}$. Such current can degrade due to electron-phonon interactions if electrons in the lower band
can be excited to the higher band. However, such excitations 
in the bands of effective Hamiltonians require a physical energy greater than the gap $\Delta$ as is rigorously
established in "Floquet Fermi golden rule" in Appendix A. 
It is in principle possible to absorb energies from photons, but because the frequency of photons $\Omega$
is assumed to be much larger than band width, the absorption of such large energy requires the 
excitations of electrons together with many phonons, and therefore 
such a process is suppressed. Thus, the transition of an electron
from the lower band of the effective Hamiltonian to the higher band is possible only through the absorption of
phonon energies. Therefore, at low temperatures, the property of an "insulating" state of the effective Hamiltonian 
is protected against electron-phonon interactions by the gap. 


\section{Probe of the induced effective gap in an isolated system} \label{sec:pumpprobe}
In this section, we propose a different way to probe the effective gap induced by off-resonant light
through the pump and probe measurements in an isolated system. 
The essential idea is simple. Given a system under the application of light (called pump laser), 
suppose that the effective
Hamiltonian $H_{\textrm{eff}}$ defined by \eqnref{defeff} has a gap $\Delta$. We prepare the 
state, in isolation from thermal reservoirs, such that only the lower band of $H_{\textrm{eff}}$ is occupied 
through a sort of "adiabatic preparation." Here we start from zero-temperature static system, 
and increase the intensity of light gradually to increase the size of the gap, $\Delta$. 
As we argue below and Appendix A, adiabatic theorem in Floquet picture guarantees that the final state has 
the electron occupations such that the lower band of photon-dressed Hamiltonian  $H_{\textrm{eff}}$ is occupied. Now for this occupation of electrons with a gap to a higher band, 
it is intuitively clear that the system becomes 
transparent to the probe light with frequency smaller than $\Delta$.
In the case of graphene and topological insulators under the application of light, 
we expect that the transmitted probe light results in the Faraday rotations\cite{}.


The pump and probe measurements described above are well-understood if the modification of 
the system from the original Hamiltonian to final Hamiltonian $H_{\textrm{eff}}$ is done through a static fields. 
In this case, the adiabatic preparation is guaranteed by adiabatic theorem, and the transmission
of probe light can be confirmed by looking at the Fermi golden rule which shows that the photons
cannot be absorbed by electrons due to the conservation of energy. 

In the case of periodically varying fields, analogous statements hold. "Floquet adiabatic theorem" 
shows that, under an adiabatic evolution of the periodically varying fields, each Floquet state 
follows the instantaneous Floquet state given by the instantaneous Hamiltonian. Similarly, "Floquet 
Fermi golden rule" gives the rate in which the transition from one Floquet state to another happens 
under small perturbations. This result shows that the quasi-energies of Floquet states are conserved
up to integer multiples of driving frequency $\Omega$. Thus as we have claimed above, the electrons 
in the lower band of $H_{\textrm{eff}}$ cannot be excited to higher bands unless the photon energy is 
larger than the band gap $\Delta$. We give the detailed proof of these theorems in the Appendix A.

\section{Conclusion} \label{sec:conclusion}
In this paper, we studied the transport properties of non-equilibrium systems under the application of light 
in many-terminal measurements. Starting from Floquet-Landauer formula, we gave two different solutions
of Floquet Green's functions that illustrate the physical origin of transport in this situation. 
We found that for generic driving frequencies, the transport involves photon-assisted conductance
and cannot be described by any static, effective Hamiltonians. 

In the case of graphene and topological insulators under the off-resonant light, the non-equilibrium transport 
does not involve photon absorptions/emissions. Rather, the electron band structures are modified through
the virtual photon absorption/emission processes. We established, through the solution of Floquet Green's function,
that such modifications are captured by the static photon-dressed Hamiltonian, 
and that the transport in this system becomes
equivalent to that described by the photon-dressed Hamiltonian. 
Remarkably, the effective Hamiltonian obtained in this way 
takes the form of Haldane model\cite{Haldane1988} with second neighbor hopping with phase accumulations 
for graphene under the application of circularly polarized light. 

One important aspect of our proposal is the opening of the gap in the photon-dressed Hamiltonian when the 
original static Hamiltonian is semi-metal and gapless. 
We gave two physical manifestations of such a gap. One is the insulating behavior of the driven system
{\it attached to the leads} (\secref{sec:insulator}). The attachment of leads is crucial to determine the electron occupation numbers. Another is the transmission of low frequency light in an isolated system(\secref{sec:pumpprobe}) after 
the adiabatic preparations of states. We argued the possibility of such pump-probe measurements by 
establishing two important extensions of well-known theorems, "Floquet adiabatic theorem" an "Floquet
Fermi golden rule" (Appendix A). 

The formalism and intuitive understanding developed in this paper can be used to study the transport
properties of a variety of systems under the application of light. It is of interests to analyze, for example, 
the transport properties of light-induced topological systems proposed in Ref. [\onlinecite{Oka2009}]. 
In addition, our analysis shows that transport under the application of light contains richer physics than
static transport. In particular, photon-assisted conductance in which electrons absorbs/emits photons during 
the propagations is the unique feature of driven systems, and 
it is interesting to analyze how such physical process results in energy conductions. 
While we focused on the response current $J_{\textrm{res}}$ in this paper, yet another aspect of 
driven systems is the presence of pump current $J_{\textrm{pump}}$ appearing in \eqnref{dccurrent}. 
It is of interests to find materials that can pump currents by simply shining light on their surface.

We thank Mark Rudner, Erez Berg, David Hsieh, Bernhard Wunsch, Shoucheng Zhang, Bertrand Halperin, Subir Sachdev, Mikhail Lukin, Jay D Sau, and Dimitry Abanin for valuable discussions.
The authors acknowledge support from a grant from the
Harvard-MIT CUA and NSF Grant No. DMR-07-05472. 
T.O. is supported by Grant-in-Aid for Young Scientists (B) 
and L.F. acknowledges the support from the Harvard Society of Fellows. 

\appendix 

\section{Floquet Fermi golden rule and Floquet adiabatic theorem}

In this section, we establish the following two statements about $H_{\textrm{eff}}$
studied in the main text; 
1) The result of many-terminal measurements of the systems under the application of light 
obtained in the main text is robust against electron-phonon interactions, as long as the energy 
of phonons dictated by the temperature of the systems is smaller than the induced gap $\Delta$;
2) The photo-induced gap $\Delta$ can be probed, in a closed system, by the transmission of 
a laser with frequency $\omega < \Delta$. 

We give arguments for the first statement by deriving an analogous theorem as Fermi golden rule 
in the periodically driven systems. When the chemical potentials of leads lie in the gap of $H_{\textrm{eff}}$, 
the non-equilibrium current in many-terminal measurements is conducted through electrons in the lower band
of $H_{\textrm{eff}}$. Such current can degrade due to electron-phonon interactions if electrons in the lower band
can be excited to the higher band. By deriving "Floquet Fermi golden rule," we demonstrate that such excitations 
in the bands of effective Hamiltonians still require a physical energy greater than the gap $\Delta$. 
It is in principle possible to absorb energies from photons, but because the frequency of photons $\Omega$
is assumed to be much larger than band width, the absorption of such large energy requires the 
excitations of electrons and many phonons, and therefore such a process is suppressed. Thus, the transition of an electron
from the lower band of the effective Hamiltonian to the higher band is possible only through the absorption of
phonon energies. Therefore, at low temperatures, the property of an "insulating" state of the effective Hamiltonian 
is protected against electron-phonon interactions by the gap. 

The proof of the second statement requires two steps. If we assume that the closed system 
with $H_{\textrm{eff}}$ can be prepared in a state such that only the lower band of $H_{\textrm{eff}}$ is occupied, 
then we can argue from the "Floquet Fermi golden rule" that the low frequency laser with $\omega < \Delta$
cannot be absorbed by the electrons. Therefore, such a system is transparent to the light. 
In order to prepare such a "filled" state of the 
effective Hamiltonian $H_{\textrm{eff}}$, we consider an adiabatic preparation. Starting from the half-filled state of 
original systems whose chemical potential lies at the Dirac points, we adiabatically increase the strength of light. 
We argue, by deriving "Floquet 
adiabatic theorem"\cite{Breuer1989}, that such procedure prepares the filled state of $H_{\textrm{eff}}$ except possibly exactly at the Dirac points. 

These two statements rely on two general theorems about periodically driven systems, 
dubbed as "Floquet Fermi golden rule" and "Floquet adiabatic theorem." 
In the following, we derive these results, using the elegant approach from "two-time" formalism\cite{Breuer1989}.
We emphasize that these results are general and have wide applications outside of what we discussed 
in this paper. 


\subsection{Two-time Schr\"odinger equation}
In order to study the dynamics of periodically driven systems, 
it is convenient to separate two time scales, a fast time scale associated with the 
driving frequency $\Omega$ and a slow time scale associated with other dynamics 
such as those of phonons. We let $t$ denote the former time scale and $\tau$ the 
latter, and obtain the Schr\"odinger equation of the slower dynamics in terms of $\tau$ 
through the replacement $i\partial/\partial t \rightarrow i\partial/\partial t + i\partial/\partial \tau$.
Then the time evolution of states for slow time scale can be written as 
\begin{eqnarray}
i \frac{\partial }{\partial \tau} \ket{\psi( \tau)}  &=& \left( \mathcal{H} +\hat{V}(\tau)  \right)  \ket{\psi( \tau)} \label{slow} \\
\mathcal{H} &=& H(t) - i\frac{\partial }{\partial t} 
\end{eqnarray}
where $H(t)$ corresponds to the Hamiltonian with periodic drives with frequency $\Omega$
and $\hat{V}(\tau) $ represents the perturbation of the system with slow frequencies compared to $\Omega$.
In the absence of the perturbation $V(\tau)$, the eigenstates of the Schr\"odinger equation above is 
given by Floquet states such that 
\begin{eqnarray}
E_{\alpha}  \ket{\Phi_{\alpha}} &=&\mathcal{H}  \ket{\Phi_{\alpha}}
\end{eqnarray}
where  $\ket{\Phi_{\alpha}(t)} = e^{iE_{\alpha} t } \ket{\phi_{\alpha}(t)} = 
\sum_{n} e^{-i\Omega n t} \ket{\phi^{n}_{\alpha}} $ is a state with a periodic structure 
$\ket{\Phi_{\alpha}(t)} = \ket{\Phi_{\alpha}(t+T)}$, and $E_{\alpha}$ is the quasi-energy of
the Floquet state, {\it i.e.} the eigenenergy of $H_{\textrm{eff}}$. Here
$\ket{\phi_{\alpha}(t)}$ represents a Floquet state which satisfies the equation 
$\mathcal{H} \ket{\phi_{\alpha}(t)} =0$. 
Note that $E_{\alpha}$ is only defined up to $\Omega$, so that physically the same Floquet state 
$\ket{\phi_{\alpha}(t)}$ in \eqnref{physicalfloquet} can be associated 
with the eigenvalue $E_{\alpha} + m \Omega$ and the state 
$\ket{\Phi^m_{\alpha}(t)} = \sum_{n} e^{-i\Omega n t} \ket{\phi^{n+m}_{\alpha}} $. 
Here we take the convention that $\ket{\Phi^m_{\alpha}(t)} $ with $m=0$ is associated with 
the quasi-energy $E_{\alpha}$ such that $-\Omega/2 \leq E_{\alpha} \leq \Omega/2$. 
The orthogonality of the eigenstates $\ket{\Phi_{\alpha}}$ can be recovered by defining the 
inner product of Floquet states as the average of the usual inner product over one period of time,
 \begin{equation}
 \doublebraket{\chi_{\alpha}|  \chi_{\beta} } \equiv \frac{1}{T} \int^{T}_{0} \braket{ \chi_{\alpha}(t) | \chi_{\beta}(t)} dt. \label{inner}
 \end{equation}
 Then we have  $ \doublebraket{\Phi_{\alpha}(t)|  \Phi_{\beta}(t) } = \delta_{\alpha, \beta}$. 

These extensions of the inner products and eigenvalue problem in periodically driven systems 
can be considered as the extension of Hilbert space to include the fast time variable $t$ as another
"spatial" variable. The inner product \eqnref{inner} in this Hilbert space integrates over $t$, 
and the time variable is now represented by the slow time variable $\tau$. 
We point out that the inner product \eqnref{inner} makes sense only when any dynamics associated with $\tau$ 
occurs in a slow time scale than the period of driving $T$. In principle, operators and states which depend on
$\tau$ change during the integration time $T$ of fast time variable $t$ due to the dependence on $\tau$.
Since we are treating $t$ and $\tau$ as independent variables, the inner product \eqnref{inner} ignores such $\tau$ dependence. As long as such changes are small, the inner product \eqnref{inner} gives a good approximation.

The crucial observation is that the slow time Schr\"odinger equation in \eqnref{slow} has the identical form as the usual 
Schr\"odinger equation, and therefore, many results for static systems can be directly extended to 
periodically driven systems through the extension of the the inner product to \eqnref{inner}. 

\subsection{Floquet Fermi golden rule}
"Floquet Fermi golden rule" gives the intuition behind 
the response of periodically driven systems under the influence of perturbations. 
In particular, the result shows that quasi-energy 
of the effective Hamiltonian $H_{\textrm{eff}}$ is a conserved quantity up to the driving frequency $\Omega$. 
In the context of our paper, this result implies electrons in the lower band of $H_{\textrm{eff}}$ cannot be excited to the upper band
if the frequencies of the perturbations, such as phonons or probe light, are smaller than the gap $\Delta$. 
Thus the transport property of a non-equilibrium system described by $H_{\textrm{eff}}$ is robust 
against phonon interactions as long as the chemical potentials of leads lie in the gap and 
phonon energies are smaller than the gap $\Delta$. 
Moreover, if one can prepare the system in the state with filled lower band of $H_{\textrm{eff}}$, then the gap
of $H_{\textrm{eff}}$ can be probed by observing the transmissions of low frequency lasers. 

This "Floquet Fermi golden rule" can be easily obtained through the "two-time" formalism 
described in the previous section. In the following, we consider the perturbations of the 
system such as phonons with frequency $\omega$ much smaller than $\Omega$ such that $\omega \ll \Omega$. 
We take the perturbation in the form $\hat{V}(\tau) = \hat{V} e^{-i \omega \tau}$. 
The usual derivation of Fermi golden rule can be applied in a straightforward fashion, and we obtain
"Floquet Fermi golden rule", which gives the rate $\gamma_{i \rightarrow f}$ 
of exciting the initial Floquet state $\ket{\phi_{i}}$
to the final Floquet state $\ket{\phi_{\textrm{f}}}$ in the presence of the perturbation $\hat{V}(\tau)$; 
\begin{equation}
\gamma_{i \rightarrow f} = \sum_{m} |\doublebraket{\Phi^{m}_{f}| \hat{V} |  \Phi^0_{i} } |^2 \delta( E_{i} + \omega - E_{f}- m\Omega).
\end{equation}
Here $E_{i}$ and $E_{f}$ are the quasi-energies of the initial and final Floquet states, respectively. 
In order to derive the result above, 
we represented the Floquet state $\ket{\phi_{i}}$ by the specific periodic state 
$\ket{\Phi^0_{i}}$. This choice is arbitrary and any other choice gives the same result.  
Since the physical Floquet state $\ket{\phi_{\textrm{f}}}$ can be represented as 
the states $\ket{\Phi^m_{f}}$ for any integers $m$, the total transition rate is given by the sum of the rate from
the state $\ket{\Phi^0_{i}}$ to states $\ket{\Phi^m_{f}}$. 

This rate has the same form as the conventional Fermi golden rule, except for the summation 
over the Floquet energy index $m$. The delta function in the equation above imposes 
the conservation of {\it quasi-energy} which is the eigenenergy of effective Hamiltonian 
$H_{\textrm{eff}}$, which means the energy is conserved up to the
driving frequency $\Omega$. This is a natural consequence of the fact that the system 
can absorb or emit the energy $\Omega$ from the periodic drives. 

From this result, it is clear that such conservation of quasi-energy prevents the excitations of 
electrons from lower band to upper band when phonon energy $\omega$ is smaller than the gap
of the system, and $\Omega$ is much larger than the total band-width of electrons. 

\subsection{Floquet adiabatic theorem}
In this subsection, we show, in analogy with the adiabatic theorem of static systems, that a Floquet state follows 
an adiabatic change of Hamiltonian and stays in the Floquet state of the instantaneous Hamiltonian. 
This result indicates that the adiabatic increase of the intensity of light can be used to 
prepare the state with filled lower band of $H_{\textrm{eff}}$, whose properties can then be probed through 
low frequency lasers as argued above. 

Starting from the slow time Schr\"odinger equation in \eqnref{slow}, we can follow the derivation of 
adiabatic theorem and prove the analogous theorem for periodically driven systems. 
Here we briefly outline the derivation. 

Suppose that the total Hamiltonian $\mathcal{H}(\tau) $ 
is slowly varying as a function of $\tau$. 
We are interested in how a Floquet state of $H(0)$ at time $\tau=0$ evolves 
under this time evolution. Let $\ket{g}$ be the initial Floquet state and $\ket{G(\tau_{0})}$ 
be the result of evolving $\ket{g}$ under $H(\tau)$ for time $\tau_{0}$.

We denote the instantaneous eigenstates of $\mathcal{H}(\tau)$ as $\ket{\alpha(\tau)}$ such hat 
$\mathcal{H}(\tau) \ket{\alpha(\tau)} = E_{\alpha}(\tau)  \ket{\alpha(\tau)} $. Then we express 
the state $\ket{G(\tau)}$ in terms of  $\ket{\alpha(\tau)} $ as 
\begin{eqnarray}
 \ket{G(\tau)} &=& \exp\left( -i \int^{\tau}_{0} E_{g}(\tau') dt' \right) \nonumber \\
 && \times \left( c_{g}(\tau) \ket{g(\tau)} + \sum_{\alpha \neq g} c_{\alpha}(\tau)   \ket{\alpha(\tau)} \right)
 \end{eqnarray}
 In the absence of degenerate states, we can solve for the coefficients $c_{\alpha}(\tau)$ in the lowest order for the slow change of Hamiltonian $\mathcal{H}(\tau)$ in the Schr\"odinger equation of \eqnref{slow}. 
 The result is given by 
 \begin{widetext}
   \begin{eqnarray}
 \ket{G(\tau)} &=&e^{ -i \int^{\tau}_{0} E_{g}(\tau') dt' } \exp\left( -i \int^{\tau}_{0}  i \doublebra{g(\tau ')} \frac{\partial}{\partial \tau'} \doubleket{g(\tau')} d\tau' \right)    \left( \ket{g(\tau)} -i \sum_{\alpha \neq g}   \ket{\alpha(\tau)} \frac{ \doublebra{\alpha(\tau)} \frac{\partial}{\partial \tau} \doubleket{g (\tau)} } {E_{\beta}(\tau)- E_{g}(\tau)}\right)
 \label{adiabatic}
 \end{eqnarray}
 \end{widetext}
 Thus to the zeroth order for the slow change of Hamiltonian $\mathcal{H}(\tau)$, 
 $ \ket{G(\tau)}$ is the Floquet state of the instantaneous Hamiltonian $\mathcal{H}(\tau)$ with possible 
 accumulations of dynamical and Berry phases. The first order correction is given by the second term 
 of \eqnref{adiabatic}. 
 
For static systems of Dirac Fermions studied in this paper, we have shown that a gap proportional to 
$\mathcal{A}^2$ opens at the Dirac point upon the application of light. If the chemical potential lies at the 
Dirac point before the application of light, the result above implies that the adiabatic increase of the 
intensity of light $\mathcal{A}(\tau)$ can be used to prepare the system close to the filled lower band state of 
$H_{\textrm{eff}}$. At exactly the Dirac points where the spectrum becomes degenerate, the adiabatic 
theorem above does not apply, but these points represent only a tiny portion of the total states, and thus 
can be ignored for the calculations of physical quantities. 
When the initial system is at finite temperature, such adiabatic increase of $\mathcal{A}(\tau)$ leads to
{\it non-thermal} distributions of electrons in the spectrum of $H_{\textrm{eff}}$, but nonetheless the 
resulting density matrix can be calculated through the result \eqnref{adiabatic} in the adiabatic limit. 

This Floquet adiabatic theorem can be used to obtain the Kubo's formula\cite{YHatsugai1997}
in the non-interacting, periodically driven systems. 
In Ref.[\onlinecite{Oka}], such result is applied to derive the extension of TKNN formula\cite{Thouless1982} to 
periodically driven systems in infinite systems.


\begin{thebibliography}{10}

%
%
%



\bibitem{Glass1974}
A.~M. Glass, D. von der Linde, \& T. J. Negran, App. Phys. Lett. {\bf 25} 233 (1974);
E. J.~H. Lee {\it et~al.}  Nat. Nano. {\bf 3} 486 (2008).

\bibitem{Xu2010}
X. Xu {\it et~al.}  Nano Lett. {\bf 10} 562 (2010).

\bibitem{Karch2010}
J. Karch {\it et~al.}  Phys. Rev. Lett. {\bf 105} 227402 (2010);
T. Hatano {\it et~al.}  Phys. Rev. Lett. {\bf 103} 103906 (2009). 

\bibitem{Miyano1997}
K. Miyano {\it et~al.}  Phys. Rev. Lett. {\bf 78} 4257 (1997);
M. Fiebig {\it et~al.}  Science {\bf 280} 1925 (1998).

  \bibitem{Syzranov2008}
S.~V. Syzranov, M. V. Fistul, \& K. B. Efetov, Phys. Rev. B {\bf 78} 045407 (2008).

\bibitem{fausti2011}
D. Fausti {\it et~al.} Science {\bf 14} 331 6014 189 (2011).

  \bibitem{Haldane1988}
F. D.~M. Haldane, Phys. Rev. Lett. {\bf 61} 2015 (1988). 
    
\bibitem{Qi2008}
X.-L. Qi, T. L. Hughes \& S.-C. Zhang, Phys. Rev. B {\bf 78} 195424 (2008);
X.-L. Qi, R. Li, J. Zang \& S.-C. Zhang, Science {\bf 323} (2009). 

\bibitem{Oka}
T. Oka \& H. Aoki, Phys. Rev. B {\bf 79} 081406 (2009);

\bibitem{Oka2009}
N.~H. Lindner, G. Rafael \& V. Galitski, Nat. Phys. {\bf 7} 490 (2011);
W. Yao, A.H. MacDonald \& Q. Niu, Phys. Rev. Lett. {\bf 99} 047401 (2007). 

\bibitem{Kitagawa2010a}
T. Kitagawa {\it et~al.}  Phys. Rev. A {\bf 82} 0033429 (2010);
T. Kitagawa {\it et~al.}  Phys. Rev. B {\bf 82} 235114 (2010);
L. Jiang {\it et~al.}  Phys. Rev. Lett {\bf 106} 220402 (2011).

\bibitem{Jauho1994}
A. P. Jauho, N. S. Wingreen \& Y. Meir, Phys. Rev. B {\bf 50} 5528 (1994);
S. Kohler, J. Lehmann \& P. Hanggi, Phys. Rep. {\bf 406} 379 (2005);
M. Moskalets \& M. B\"{u}ttiker, Phys. Rev. B {\bf 66} 205320 (2002);
L. Arrachea \& M. Moskalets, Phys. Rev. B {\bf 74} 245322 (2006);
Hernan L. Calvo, {\it et al}, Appl. Phys. Lett. {\bf 98}, 232103 (2011);
D. Martinez, R. Molina \& B. Hu, {\bf 78} 045428 (2008).

\bibitem{Thouless1982}
D. Thouless {\it et~al.}  Phys. Rev. Lett. {\bf 49} 405 (1982).

\bibitem{Buttiker1986}
M. B\"{u}ttiker, Phys. Rev. Lett. {\bf 57} 1761 (1986). 

\bibitem{sambe}
H. Sambe, Phys. Rev. A {\bf 7}, 2203 (1973),
M. Torres \& A. Kunold, Phys. Rev. B {\bf 71}, 115313 (2005).



\bibitem{Switkes1999}
M. Switkes {\it et~al.} Science {\bf 283} 1905 (1999);
L. E. F. FoaTorres, Phys. Rev. B {\bf 72} 245339 (2005). 

\bibitem{Martinez2003}
D. F. Martinez, J. Phys. A: Math. Gen. {\bf 36} 9827 (2003),
T. Brandes and J. Robinson, Phys. Status Solidi B {\bf 234}, 378, (2002),
D. F. Martinez and R. A. Molina, Eur. Phys. J. B {\bf 52}, 281 (2006).

\bibitem{inoue2010}
J-i. Inoue \& A. Tanaka, Phys. Rev. Lett {\bf 105} 017401 (2010). 

\bibitem{Schoenlein2000}
R.~W. Schoenlein  \emph{et~al.}, Science {\bf 287} 2237 (2000).

\bibitem{Zhang2009}
H. Zhang {\it et~al.} Nat. Phys. {\bf 5} 438 (2009);
Y. Xia {\it et~al.} Nat. Phys. {\bf 5} 398 (2009). 


\bibitem{Fu2007}
L. Fu \& C. L. Kane, Phys. Rev. B {\bf 76} 045302 (2007).

\bibitem{Okafaraday}
T. Oka \& H. Aoki arXiv:1007.5399 (2010).

\bibitem{Novoselov2005}
K.~S. Kovoselov \emph{et~al.} Nature {\bf 438} 197 (2005). 

\bibitem{Breuer1989}
\bibinfo{author}{Breuer, H.}
\newblock \bibinfo{title}{{Quantum phases and Landau-Zener transitions in
  oscillating fields}}.
\newblock \emph{\bibinfo{journal}{Physics Letters A}}
  \textbf{\bibinfo{volume}{140}}, \bibinfo{pages}{507--512}
  (\bibinfo{year}{1989}),
  S. C. Althorpe, D. J. Kouri, D. K. Hoffman, and N. Moiseyev, Chem. Phys. {\bf 217},
289 (1997).
  
  

\bibitem{YHatsugai1997}
\bibinfo{author}{{Hatsugai,Y}}.
\newblock \bibinfo{title}{{Topological aspects of the quantum Hall effect}}
\newblock \emph{\bibinfo{journal}{Journal of Physics: Condensed Matter}}
  \textbf{\bibinfo{volume}{9}}, \bibinfo{pages}{2507}
  (\bibinfo{year}{1997}).
  

\bibitem{Thouless1982}
D. Thouless {\it et~al.}  Phys. Rev. Lett. {\bf 49} 405 (1982).


\end{thebibliography}
\end{document}